\documentclass[aps,prb,reprint]{revtex4-2}
\usepackage{amsmath}
\usepackage{amssymb}
\usepackage{bm}
\usepackage{color}
\usepackage[utf8]{inputenc}
\usepackage{pifont}
\usepackage[colorlinks,linkcolor={blue},citecolor={blue},urlcolor={blue}]{hyperref}
\usepackage{mathtools}
\usepackage{booktabs}
\newcommand{\Top}{{\rm T}}
\newcommand{\Bot}{{\rm B}}

\newcommand{\bk}{{\bm k}}
\newcommand{\bq}{{\bm q}}

\newcommand{\zh}{\hat{z}}

\newcommand{\bkpar}{{\bm k}_{ \scriptscriptstyle \parallel}}
\newcommand{\bqpar}{{\bm q}_{ \scriptscriptstyle \parallel}}

\newcommand{\kpar}{ k_{ \scriptscriptstyle \parallel}}
\newcommand{\sbkpar}{{ \bm k}_{{\tiny{\text{$ \scriptscriptstyle \parallel$}}}}}

\newcommand{\bsigma}{{\bm \sigma}}
\newcommand{\bqz}{{\bm q}_{z}}

\begin{document}

\title{Photo-induced Non-collinear Interlayer RKKY Coupling in Bulk Rashba Semiconductors}

\author{Mahmoud M. Asmar}
\affiliation{Department of Physics, Kennesaw State University, Marietta, Georgia 30060, USA}
\author{Wang-Kong Tse }
\affiliation{Department of Physics and Astronomy, The University of Alabama, Tuscaloosa, AL 35487, USA }
\date\today

\begin{abstract}
The interplay between light-matter, spin-orbit, and magnetic interactions allows the investigation of light-induced magnetic phenomena that are otherwise absent without irradiation. We present our analysis of light-driven effects on the interlayer exchange coupling mediated by a bulk Rashba semiconductor in a magnetic multilayer.
The collinear magnetic exchange coupling mediated by the photon-dressed spin-orbit coupled electrons of BiTeI develops light-induced oscillation periods and displays new decay power laws, both of which are enhanced with an increasing light-matter coupling. For magnetic layers with non-collinear magnetization, we find a non-collinear magnetic exchange coupling uniquely generated by light-driving of the multilayer. As the non-collinear magnetic exchange coupling mediated by the photon-dressed electrons of BiTeI is unique to the irradiated system and it is enhanced with increasing light-matter coupling, this effect offers a promising platform of investigation of light-driven effects on magnetic phenomena in spin-orbit coupled systems. In this platform, light properties, such as its intensity, can serve as external knobs for inducing non-collinear couplings of the interlayer exchange and for modulating the collinear couplings. Both of these effects signify the photo-generated modification in the spin textures of spin-orbit coupled electrons in BiTeI.
\end{abstract}

\maketitle

\section{Introduction}
The control of material properties and phases via dynamic drives has recently gained a burgeoning interest. Specifically, the modulation of matter by optical drives has been greatly aided by the state-of-the-art development of stable and high-intensity sources of radiation that operate at a broad range of wavelengths~\cite{light0,light2}. The engineering of systems' properties via periodic driving, i.e., Floquet engineering~\cite{flreview1,flreview2,flreview3,Oka_RMP}, has motivated the theoretical investigation of light-controlled and generated phenomena and phases in matter, such as transport properties~\cite{Platero,ftrans1,
Floqtrans3,martin1,Asmar2022}, topological phases of matter~\cite{FloqTIReview,Floqtop3,Floqtop2,sentef1,opt2,opt1,Torres_graphene,sentef2,martin3,Bicircular}, magnetic exchange interactions~\cite{Asmar2020,Asmar2021}, spin-injection~\cite{spininj}, and tunneling phenomena~\cite{FloqTunn4,FloqTunn1}. Experimentally, light-induced topological transitions were detected via optical conductivity measurements in graphene systems~\cite{FloqExp3}, and photon-dressed surface bands of topological insulators have been observed by time-resolved and angle-resolved photoemission spectroscopy~\cite{FloqExp1,FloqExp2}.

The indirect interaction between magnetic adatoms embedded in metals or ferromagnets in heterostructures with non-magnetic spacers is mediated by the conduction electrons of the non-magnetic material, i.e., the Ruderman–Kittel–Kasuya–Yosida (RKKY) interaction.
The RKKY exchange coupling oscillates with the embedded magnetic impurities separation and with spacer thickness in magnetic heterostructures. In both of these cases, the oscillation period is set by the Fermi wavelength of the non-magnetic material~\cite{kittelbook}. The oscillation envelope decays as a power-law determined by the dimensionality of the non-magnetic material and the nature of its electrons~\cite{Graphenerkky,RKKYintgraph,MOS2RKKY}.
Conventionally, control of the RKKY interaction in materials and heterostructures has been achieved by static means,
such as gate voltage variations~\cite{RKKYvolt3,RKKYvolt4}, control of the non-magnetic layer thickness in heterostructures~\cite{exp1,exp2,LayersRKKY1}, or implementation of low dimensional materials with exotic fermions~\cite{Graphenerkky,RKKYintgraph,MOS2RKKY}.

Light-driving of materials leads to the photon-dressing of their Bloch bands and the formation of Floquet-Bloch states~\cite{flreview2,flreview3,flreview4,Oka_RMP}. Light's frequency and intensity can tune these states that display properties not present in their parent equilibrium system. Hence, photon-dressing of electrons in light-driven materials is key to dynamically controlling the magnetic exchange coupling. However, it has been shown in Refs.~\cite{Asmar2021,Asmar2020,ono} that the effects of periodic monochromatic light-driving on the magnetic exchange interaction mediated by two dimensional (2D) spin-degenerate systems, such as 2D electron gases (2DEGs) and graphene, are limited to light-controlled collinear exchange couplings. The main effects of light on the magnetic exchange mediated by these spin-degenerate systems are displayed through a light-controlled Fermi wavevector that leads to photo-controlled RKKY oscillations, and the “draining out” of the Fermi sea, either by an increasing photoinduced gap (for graphene) or by the migration of the zeroth Floquet band (for 2DEGs) above the Fermi level, which lead to a non-oscillatory behaviour of the exchange coupling in what resembles the Bloembergen-Rowland interaction~\cite{BR}. On the other hand, 3D and 2D spin-orbit coupled materials offer a rich platform to explore the interplay between magnetic and spin-orbit effects to provide the basis for wide-ranging spintronics phenomena~\cite{Spintronics1}. These systems also allow for the exploration of optical induction and control of non-collinear magnetic couplings which play an essential role in the formation of magnetic skyrmions~\cite{skyrmions}, spin helices~\cite{helix}, and chiral domain walls~\cite{domainwall}.

Here, we show that a non-collinear magnetic exchange [Dzyaloshinskii-Moriya (DM)]~\cite{DM1,DM2} coupling can be photo-generated in systems that otherwise lack this interaction at equilibrium. We consider a monochromatically irradiated magnetic heterostructure composed of 2D ferromagnets~\cite{EuS2,EuS3,EuS4}, enclosing the 3D Rashba semiconductor BiTeI.
Our analysis of the irradiated multilayer system reveals the photon-dressing of BiTeI's electronic bands and the modification of their spin content. Consequently, the determination of the irradiated system's spin susceptibility tensor and the interlayer magnetic exchange coupling components shows that the light-induced change in the electronic bands' spin textures relaxes the electrons scattering constraints leading to light-induced non-collinear interlayer magnetic exchange interactions.

\section{Model}
In order to study the effects of periodic driving on the interlayer magnetic exchange interaction mediated by spin-orbit active materials we consider a magnetic multilayer composed of BiTeI sandwiched by top and bottom ferromagnets deposited along its stacking direction, as shown in Fig.~\ref{fig1}(a). Then we subject the multilayer to a monochromatic and circularly polarized light.

\subsection{Rashba Semiconductor, BiTeI}
\begin{figure}
  \centering
  \includegraphics[width=\columnwidth]{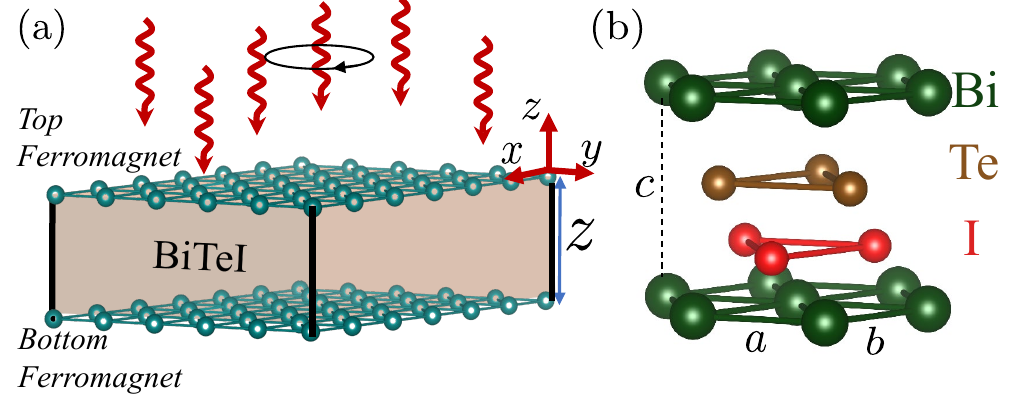}
  \caption{(a) Schematic representation of an irradiated Rashba magnetic multilayer. The multilayer is composed of two ferromagnetic layers enclosing BiTeI. The distance between the ferromagnetic layers is $z$ and the incoming circularly polarized light is normally incident to the $x-y$ plane. (b) Crystal structure of BiTeI. The axes $a$, $b$, and $c$ indicate the crystallographic axes.}\label{fig1}
\end{figure}

\begin{figure*}
  \centering
  \includegraphics[width=\textwidth]{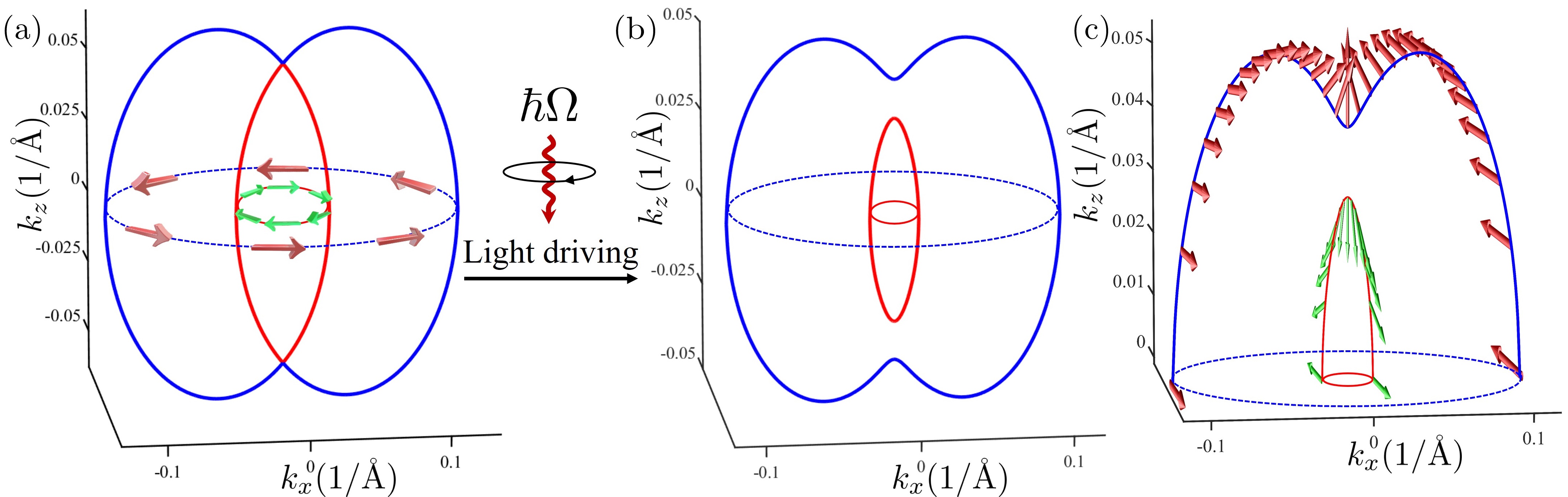}
  \caption{(a) Fermi surface of BiTeI before irradiation for $E_F>0$. In this case the inner and outer portions of the Fermi surface host states with definite and opposite helicities that do not change with $k_z$. (b) BiTeI's modified Fermi surface shape due to photon-dressing of the bands. (c) Modified spin structure of states at the Fermi surface under light-driving. Unlike the non-irradiated case in (a), the light-driven case in (c) shows variations of the spin components with $k_z$.}\label{fig2}
\end{figure*}

Similar to BiTeCl and BiTeBr, BiTeI has a layered lattice structure which belongs to a family of semiconductors known as Rashba semiconductors~\cite{BiTeI1,BiTeI2,BiTeI3,BiTeI6}. BiTeI crystal structure has an intrinsically broken inversion symmetry along its stacking, $c$, axis due to its asymmetrically stacked triangular layers of Bi, Te and I, as shown in Fig.~\ref{fig1}(b). The lack of inversion symmetry along the $c$-axis causes the electrons in the $a$-$b$ plane to experience a spin-orbit interaction that inherits the three-fold symmetry of this plane. Therefore, the intrinsic spin-orbit interaction in BiTeI takes the form of the Rashba spin-orbit coupling (SOC), $\alpha_{\rm R}(\bsigma\times \bk)\cdot{\zh}$, where  $\bsigma=(\sigma_{x},\sigma_{y},\sigma_{z})$ is the vector of Pauli
matrices representing spins, $\bk=(k_{x},k_{y},k_{z})$, and $\alpha_{\rm R}$ is the Rashba SOC strength.
The low-energy quasiparticle excitations in BiTeI reside at the Brillouin zone's hexagonal face center known as the A-point [A$=(0,0,\pm\pi/c)$]~\cite{TransportBiTeI,OpticalBiTeI}, and are described by the effective Hamiltonian
\begin{equation}\label{effective1}
H_{{\rm BiTeI}}(\bk)=Ak_{z}^{2}+B\bkpar^{2}+\alpha_{\rm R}(\bsigma\times \bk)\cdot {\zh}\;.
\end{equation}
Here, $A=\hbar^2/(2m_z)\approx 8.04$ eV$\rm\AA^{2}$, $B=\hbar^2/(2m_\parallel)\approx 40.21$ eV$\rm\AA^{2}$, $m_z$ and $m_\parallel$ are the effective mass tensor components, and the Rashba SOC strength $\alpha_{\rm R}\approx 3.85$ eV$\rm\AA$, $c= 6.854\;\rm\AA$ and $a=4.34\;\rm\AA$ have been reported in the literature from transport and optical experiments, and density functional theory~\cite{BiTeI1,BiTeI2,BiTeI3,BiTeI6,TransportBiTeI,OpticalBiTeI}. The low-energy approximation for the A-point electrons of BiTeI,
Eq.~\eqref{effective1}, is valid up to a cutoff energy $E_c \approx 0.2$ eV. Beyond this approximation BiTeI electronic bands acquire trigonal warping effects and additional bands not captured by this approximation. Stoichiometric BiTeI is an n-doped semiconductor with its Fermi level above the Dirac node resulting from the Rashba SOC~\cite{BiTeI1,BiTeI2,TransportBiTeI}. For $E_{\rm F}>0$ the Fermi surface is composed of two portions, as shown in Fig.~\ref{fig2}(a). The low-energy states in BiTeI are distinguished by their helicity since $[H_{{\rm BiTeI}},\hat{h}]=0$, where $\hat{h}=(\bsigma\times \bk)\cdot{\zh}/\kpar$ is the helicity operator. Moreover, each section of the Fermi surface hosts states with a unique helicity. At each segment of the Fermi surface the helical states take the form $\vert \bm{k},\mu\rangle=\left(\begin{array}{cc}i &\mu e^{i\theta_{\bk}} \\ \end{array} \right)^{\rm T}e^{i(\sbkpar\cdot{\bm r}+k_{z}z)}$, where $\mu=\pm$ is the helicity of the eigenstate, $\theta_{\bk}=\tan^{-1}(k_{y}/k_{x})$ is the azimuthal angle of the in-plane momentum $\bkpar=(k_{x},k_{y},0)$. Moreover,  the Fermi surface states at each segment have unique spin textures, where $\langle\sigma_{x}\rangle=\mu\sin\theta_{\bk}$, and $\langle\sigma_{y}\rangle=-\mu\cos\theta_{\bk}$, while $\langle\sigma_{z}\rangle=0$, as shown in Fig.~\ref{fig2}(a).  Hence, at equilibrium, the presence of the Rashba SOC leads to the coupling of the in-plane momentum of the low-energy excitations in BiTeI and their spins, which makes these states helical with spins restricted to the $x-y$ ($a-b$) plane, as shown in Fig.~\ref{fig2}(a).

\subsection{Irradiated Rashba Semiconductor}
When the driving field is initially activated, energy begins to be pumped into the system. The simultaneous presence of energy injection and relaxation through various inelastic scattering processes (such as electron-electron and electron-phonon scattering) does not allow for indefinite heating~\cite{Asmar2020}. After initial transients subside, which also include the buildup time for the RKKY interactions~\cite{build1,build2} the system dynamics settle into a nonequilibrium steady state (NESS), where time periodicity is restored. Moreover, when a high-frequency off-resonant light is illuminated on a material, its electrons cannot directly absorb photons~\cite{note1}. Therefore, the electron dynamics of the system are well described by an approximate static Hamiltonian that captures the virtual photon absorption. While subjecting the Rashba magnetic multilayer to an off-resonant light we need to ensure that light penetrates our system homogeneously, and this requires a careful choice of the light frequency ($\Omega$), the Rashba semiconductor thickness ($z$), and the thin ferromagnets enclosing BiTeI. Based on the band structure of BiTeI ~\cite{BiTeI1,BiTeI2,BiTeI3,BiTeI6}, we have deduced that the low-energy quasiparticle excitations at the A point experience an off-resonant light-driving for light frequencies in the range $\hbar\Omega\geq 7.5$ eV. Moreover, the sample thickness should satisfy $z\ll \delta$, where $\delta$ is the skin depth which depends on the light's frequency $\delta\propto\Omega^{-1/2}$. The skin depth of midinfrared light on BiTeI has been experimentally measured, $\delta\sim 10–30\;\mu$m~\cite{OpticalBiTeI}. Then, from these values of $\delta$ we can deduce that the thickness of the BiTeI sample can be in the range of $z\sim 0.05-0.5\;\mu$m for our driving frequency. In addition to the constraints on the frequency and sample thickness, we need to consider insulating ferromagnetic layers sandwiching BiTeI in order to increase the amount of light that penetrates the Rashba magnetic multilayer.

At equilibrium, the magnetic exchange interaction between the two ferromagnetic layers composing the multilayer system in Fig.~\ref{fig1}(a) is mediated by the low-energy electrons of BiTeI. On the other hand, an irradiated magnetic multilayer with thin ferromagnets and $z\ll \delta$, will have an indirect magnetic exchange interaction that is mediated by the photon-dressed electrons of BiTeI, since the entirety of the BiTeI sample is homogeneously irradiated, its electrons experience a uniform light-matter interaction with the driving field. This leads to the formation of photon-dressed bands, with the photon-dressed electrons acting as the  effective quasiparticles in the system that mediate the RKKY interaction between the ferromagnetic layers. To capture the effects of irradiation on the low-energy BiTeI quasiparticles, we consider a monochromatic light that is circularly polarized and normally incident to the $x-y$ plane of the multilayer, as shown in  Fig.~\ref{fig1}(a). The time-periodic BiTeI Hamiltonian follows from the minimal coupling $\bk\rightarrow \bk+(e/\hbar){\bm A(t)}$ in $H_{{\rm BiTeI}}$ [Eq.~\eqref{effective1}], where ${\bm A(t)}=\sqrt{2}A_0{\rm Re}[ie^{-i\Omega t}{\bm \hat{e}_\tau}]$, ${\bm \hat{e}_\tau}=(\hat{x}-i\tau\hat{y})/\sqrt{2}$, $\tau=\pm1$ indicates left and right circular polarization, $A_0=E_0/\Omega$, $E_0$ is the electric field amplitude of the incoming light, $\Omega$ is the light's frequency, and $e>0$ is the electron charge. Time-periodicity of the light-matter coupled BiTeI Hamiltonian allows the use of Floquet theory~\cite{Shirley} and the determination of the full Floquet Hamiltonian,
\begin{equation}\label{FloquetH}
H^{\rm F}_{m,n}=(H_0-n\hbar\Omega)\delta_{m,n}+H_{-1}\delta_{m+1,n}+H_{1}\delta_{m-1,n}\;.
\end{equation}
Here, $H_0(\bk)=H_{{\rm BiTeI}}(\bk)+\mathcal{V}$, where the light-induced potential shift $\mathcal{V}=B\mathcal{A}^2$, $\mathcal{A}=eA_0/\hbar$, $H_{\pm1}=\mathcal{A}(\pm B k_{\mp}+\alpha_{\rm R}{\sigma_{\pm}})$, $k_{\pm}=k_{x}\pm ik_{y}$, $\sigma_{\pm}=(\sigma_x\pm i\sigma_y)/2$, and $\tau=1$ henceforward. For an off-resonant driving field, the effects of irradiation can be described by an effective Floquet Hamiltonian that is derived from the van Vleck perturbation theory~\cite{VanVleck,martin2}. This approach utilizes the blocks of the full Floquet Hamiltonian and allows us to find the light-induced terms in powers of $1/\Omega$. By retaining terms up to order $1/\Omega$, we have
\begin{equation}\label{vVeffe}
  \mathcal{H}(\bk)=H_{0}(\bk)+\sum_{n>0}\frac{[H_{n},H_{-n}]}{n\hbar\Omega}+\mathcal{O}(\Omega^{-2})\;.
\end{equation}
For our system the index $n=0,\pm1$ due to the block-tridiagonal nature of the system's full Floquet Hamiltonian, and $H_{0}$ and $H_{\pm1}$ are given in Eq.~\eqref{FloquetH}. Substituting $H_{n=0,\pm1}$ in Eq.~\eqref{vVeffe}, we obtain the effective Floquet Hamiltonian for off-resonantly irradiated BiTeI,
\begin{equation}\label{HeffSys}
\mathcal{H}(\bk)=Ak_{z}^{2}+B\bkpar^{2}+\mathcal{V}+\alpha_{\rm R}(\bsigma\times \bk)\cdot{\zh}+\Delta\sigma_{z}\;,
\end{equation}
where $\Delta=(\mathcal{A}\alpha_{\rm R})^2/\hbar\Omega$, $\mathcal{A}$ and $\mathcal{V}$ are given in Eq.~\eqref{FloquetH}. From the effective Floquet Hamiltonian one can notice that the principal effects of off-resonant irradiation of BiTeI are the generation of the mass term $\Delta$ and the generation of a light-induced energy shift $\mathcal{V}$. These effects are captured in the photon-dressed electron's energy spectrum, i.e,
\begin{equation}\label{energies}
 E_{\bk,\mu}=Ak^{2}_z+B\bkpar^2+\mu\epsilon_{\bkpar}+\mathcal{V}\;,
\end{equation}
where $\mu=\pm1$, $\epsilon_{\bkpar}=\sqrt{(\alpha_{\rm R}\bkpar)^2+\Delta^2}$. Here, we should point out that within the off-resonant high-frequency regime our choice of circularly polarized light ensures the opening of a maximally-sized light-generated gap. This is because the gap ranges from zero for linearly polarized light to maximum for circularly polarized light~\cite{FloqTIReview}, so for elliptically polarized light the size of the gap thus created is always smaller than the one generated by circularly polarized light.
In addition to the modification of the energy spectrum, off-resonant driving also alters the spin properties of the electronic states. The corresponding photon-dressed electronic eigenstates of the system become
\begin{equation}\label{basis}
|\bk,\mu\rangle=\left(
                  \begin{array}{c}
                    \cos\left(\frac{\phi_{\bk}}{2}-\frac{(1-\mu)\pi}{4}\right) \\
                    -ie^{i\theta_{\bk}}\sin\left(\frac{\phi_{\bk}}{2}-\frac{(1-\mu)\pi}{4}\right)\\
                  \end{array}
                \right)e^{i\bkpar\cdot{\bm r}}e^{ik_{z}z}\;,
\end{equation}
here we have defined $\cos(\phi_{\bk})=\Delta/\epsilon_{\bkpar}$, $\sin(\phi_{\bk})=k\alpha_{\rm R}/\epsilon_{\bkpar}$, and $\tan(\theta_\bk)=k_y/k_x$.

In the off-resonant high-frequency driving regime, hybridization between different Floquet bands is suppressed, and the occupation of the Floquet bands is well described by the Fermi-Dirac distribution~\cite{ftrans1,off-res}. The spectral properties of the non-irradiated and irradiated systems are captured in the Fermi surfaces in Fig.~\ref{fig2}(a) and (b). Similar to the Fermi surface of BiTeI at equilibrium, irradiated BiTeI's displays a Fermi surface with inner and outer portions, however, the Dirac's nodes degeneracy present at equilibrium is lifted by irradiating the Rashba semiconductor.  Notably, the effects of light on BiTeI are not limited to the photon-dressing of its energy spectrum and Fermi surface but also extend to its states’ spin properties. The spin textures of the irradiated photon-dressed states become,
\begin{eqnarray}\label{texture}
&&\langle\sigma_x\rangle=\mu \sin(\phi_{\bk})\sin(\theta_{\bk}),\;\langle\sigma_y\rangle=-\mu\sin(\phi_{\bk})\cos(\theta_{\bk}),\nonumber\\
&&{\rm and}\;\langle\sigma_z\rangle=\mu\cos(\phi_{\bk}).
\end{eqnarray}
Moreover, as shown in Fig.~\ref{fig2}(c), the photon-dressed states become non-helical and acquire out-of-plane spin components that vary across the $k_z$ direction of the Fermi surface. This is unlike the non-irradiated case which hosts states with non-varying helicities along the Fermi surface's $k_z$ axis.

\section{Formulation of the Interlayer Exchange Interaction}~\label{sec2}
In the system in Fig.~\ref{fig1}(a), the irradiated Rashba semiconductor BiTeI acts as the spacer separating the top and bottom ferromagnets ($F_{\Top}$ and $F_{\Bot}$) deposited along BiTeI's stacking direction. The separation between the two ferromagnetic layers $z$ is an integer multiple of the BiTeI unit cell thickness $c$, {\it i.e.}, $z=(N+1)c$. The ferromagnetic layers neighboring BiTeI consist of classical spins $\bm S_{i}$ that locally couple to the BiTeI electrons spins through an interfacial contact potential, $\mathcal{V}_{i}=J_{0}\delta(\bm r-\bm R_{i})\mathcal{S}\cdot\bm S_{i}$, where $\bm R_{i}$ are the atomic positions at the interface of BiTeI, $\mathcal{S}$ is the spin operator of its electrons, and $J_{0}$ is the amplitude of the potential. Since every spin of the ferromagnetic layer $\bm S_{i}$ must be located on an atomic location $\bm R_{i}$ at the surface of BiTeI, this constrains the ferromagnetic layers to have a lattice structure commensurate with the terminated surfaces of BiTeI. The adoption of these considerations enables us to write the expression of the interlayer exchange as
\begin{eqnarray}\label{internonsim}
&&{\rm J}(z)=-\sum_{\alpha,\beta=x,y,z}\frac{J^{2}_{0}S^{\Top}_{\alpha}S^{\Bot}_{\beta}c}{2\mu^2_{\rm B}(2\pi)^{3}V_{0}}\\&&\times\int_{-{\pi}/{c}}^{{\pi}/{c}}dq_{z}e^{iq_z z}\nonumber \int_{{\rm2DBZ}}d^{2}{\bqpar}\chi_{\alpha \beta}(\bqpar,q_{z}) \sum_{\bm R\in  F_{{\rm T}}}e^{i\bqpar\cdot \bm R}\;.\nonumber
\end{eqnarray}
Here we have defined the spin projections of the
top (T) and bottom (B) ferromagnetic layers as $S^{(\Top,\Bot)}_{x,y,z}$, $V_{0}$ is the unit-cell volume, and $\chi_{\alpha \beta}(\bqpar,q_{z})$ is the $\alpha \beta$
component of the static spin susceptibility tensor. The dimensions of the $x-y$ plane of the system, in Fig.~\ref{fig1}(a), are much larger than the distance between the ferromagnets. We assume periodic boundary conditions along the $x$ and $y$ directions, and hence the interlayer exchange becomes
\begin{eqnarray}\label{internonsim1}
&&{\rm J}_{\alpha\beta}(z)=\\&&-\frac{1}{2}\left(\frac{J_{0}}{\mu_{\rm B}V_{0}}\right)^2\frac{S^{\Top}_{\alpha}S^{\Bot}_{\beta}c^{2}}{2\pi }\int_{-\pi/c}^{\pi/c}{dq_{z}} e^{iq_{z}z}\chi_{\alpha\beta}({\bqpar}=0,q_{z}).\nonumber
\end{eqnarray}
To obtain the latter expression we used the fact that the last sum in Eq.~\eqref{internonsim} is nonzero only for $\bqpar=0$, and that the area of the projected 2D BZ for BiTeI is $(2\pi)^{3}c/(2\pi
V_{0})$.

The components of the spin susceptibility, $\chi_{\alpha \beta}$, for BiTeI
consist of intraband and interband contributions. The non-interacting spin susceptibility of a
spin-orbit-coupled material can be obtained from the Matsubara formalism, such that the spin susceptibility can be written as
\begin{eqnarray}\label{greensusc}
&&\chi_{\alpha \beta}(\bq,iq_{n})=\\&&-\mu^{2}_{B}k_{B}T\sum_{ik_{n}}\sum_{\bk}{\rm Tr}\left\{G_{\bk}(ik_{n})\sigma_{\alpha} G_{\bk+\bq}(ik_{n}+q_{n})\sigma_{\beta}\right\}\;,\nonumber
\end{eqnarray}
where $\mu_{B}$ is the Bohr magneton, $k_{n}$ and $q_{n}$ are the fermionic Matsubara
frequencies, Tr denotes a trace, $k_{B}$ is the Boltzmann constant, and $T$ is
the temperature. The Matsubara Green's function $G_{\bk}(ik_{n})$ can be obtained from
\begin{equation}\label{fmatsubara}
  G_{\bk}(ik_{n})=\sum_{\mu=\pm}\frac{|\bk,\mu\rangle\langle\mu,\bk|}{ik_{n}-E_{\bk,\mu}}\;,
\end{equation}
where $|\bk,\mu\rangle$ is the spinor wavefunction of the $\mu^{\mathrm{th}}$ energy-band eigenstate, and $\mu = \pm$.  Substituting
Eq.~\eqref{fmatsubara} into Eq.~\eqref{greensusc}, then using the invariance of the trace under cyclic permutations, performing the Matsubara sum, and analytically continuing to the real frequency
$iq_{n}\rightarrow\omega+i\delta$ ($\delta$ is a positive infinitesimal), one obtains the expression of the retarded spin susceptibility
\begin{eqnarray}\label{spinsucesfeqfinal}
  &&\chi_{\alpha \beta}(\bq,\omega)=\\&&-\mu^{2}_{B}\sum_{\bk}\sum_{\mu,\nu}\frac{f(E_{\bk,\mu})-f(E_{\bk+\bq,\nu})}{E_{\bk,\mu}-E_{\bk+\bq,\nu}+\omega+i\delta} \mathcal{F}^{\mu\nu}_{\alpha\beta}(\bk,\bk+\bq).\nonumber
\end{eqnarray}
Here, $f(E_{\bk,\mu})$ is the Fermi-Dirac distribution function, and $\mathcal{F}^{\mu\nu}_{\alpha\beta}(\bk,\bk+\bq)=\langle
\mu,\bk |\sigma_{\alpha}|\bk+\bq, \nu\rangle \langle
\nu,\bk+\bq|\sigma_{\beta}|\bk, \mu\rangle$ is a form factor. The determination of the interlayer exchange interaction in Eq.~\eqref{spinsucesfeqfinal} requires the spin susceptibility in its static limit,
$\omega\rightarrow0$, and with $\bqpar=0$. For compactness, henceforth we omit the $\bqpar=0$ argument in $\chi_{\alpha\beta}$ and denote
$\chi_{\alpha\beta}(q_{z}) \equiv \chi_{\alpha}(\bqpar=0,q_{z})$, such that in our system $\chi_{\alpha\beta} = \sum_{\mu,\nu}\chi_{\alpha\beta}^{\mu\nu}$ with
\begin{eqnarray}\label{chispin}
\chi^{\mu\nu}_{\alpha\beta}(q_{z})=&& \frac{-\mu^{2}_{B}}{(2\pi)^{3}}\int_{-\pi/c}^{\pi/c}dk_{z} \int_{{\rm2DBZ}}{ d^{2}\bkpar }\\ &&\times\frac{f(E_{\bk,\mu})-f(E_{\bk+q_{z},\nu})}{E_{\bk,\mu}-E_{\bk+\bqz,\nu}+i\delta}\mathcal{F}^{\mu\nu}_{\alpha\beta}(\bk,\bk+\bqz).\nonumber
\end{eqnarray}

\section{Results and Discussion}
The oscillatory nature of the RKKY-mediated exchange coupling results from the spin density oscillations generated by the interaction with the ferromagnets. These oscillations are characterized by a period dictated by the wavevectors connecting to the extremal points on the Fermi surface (nesting vectors). In the low-energy regime where the quasiparticle description in Eq.~\eqref{HeffSys} is valid, the largest nesting vectors traversing the interior of the Fermi surface, Fig.~\ref{fig3}, are much smaller than $\pi/c$, and they lead to observable periods. However, the nesting vectors spanning the extremal points on the Fermi surface from its exterior, even if allowed, lead to periods smaller than $2c$ which is the smallest observable period~\cite{stiles}. Hence, we can obtain a complete description of the interlayer exchange coupling in our system by extending the $k_z$ and $q_z$ integration limits in Eqs.~\eqref{internonsim1} and \eqref{chispin} to $(-\infty$$,$$\infty)$~\cite{Asmar2019}. In the our calculations, we 
assume an ideal occupation of the Floquet bands in Eq.~\ref{energies} up to the Fermi level~\cite{Ideal1,Ideal2,Ideal3,Ideal4,Ideal5,Ideal6,Ideal7}, $E_F=180$ meV. We also note that the integration limits of Eq.~\eqref{chispin} are determined by the Fermi function and restricted to the boundaries of the Fermi surface:
\begin{eqnarray}\label{chispin2}
&&\chi^{\mu\nu}_{\alpha\beta}(q_{z}) =\frac{-\mu^{2}_{B}}{(2\pi)^{3}}\\ &&\int dk_{z}\,\int{ d^{2}\bkpar }\frac{f(E_{\bk,\mu})-f(E_{\bk+\bqz,\nu})}{E_{\bk,\mu}-E_{\bk+\bqz,\nu}+i\delta}\mathcal{F}^{\mu\nu}_{\alpha\beta}(\bk,\bk+\bqz)\nonumber\;.
\end{eqnarray}
Moreover, as the spinors in Eq.~\eqref{basis} are dependent on the momentum $q_z$ through $e^{i k_z z}$, the form factors $\mathcal{F}^{\mu\nu}_{\alpha\beta}$ are independent of the
momentum along $z$ and are given in Table~\ref{table}.
\begin{table} 
  \centering
\begin{tabular}{||c|c |c ||}
  \hline \hline
    $\alpha\beta$ & Intraband $\mu=\nu$ & Interband $\mu\ne\nu$\\ [0.5ex]
  \hline\hline
   $z z$  & $\cos^{2}\phi_\bk$                & $\sin^{2}\phi_\bk$ \\
   $x x$  & $(\sin\phi_\bk\sin\theta_\bk)^2$  & $\cos^{2}\theta_\bk+(\cos\phi_\bk\sin\theta_\bk)^2$ \\
   $y y$  & $(\sin\phi_\bk\cos\theta_\bk)^2$  & $\sin^{2}\theta_\bk+(\cos\phi_\bk\cos\theta_\bk)^2$ \\
   $z x$  & $(\sin\theta_\bk\sin2\phi_\bk)/2$   & $-i\mu \cos\theta_\bk\sin\phi_\bk-(\sin\theta_\bk\sin2\phi_\bk)/2$\\
   $z y$  & $-(\cos\theta_\bk\sin2\phi_\bk)/2$  & $-i\mu \sin\theta_\bk\sin\phi_\bk+(\cos\theta_\bk\sin2\phi_\bk)/2$\\
   $x y$  & $-(\sin2\theta_\bk\sin^{2}\phi_\bk)/2$& $i\mu \cos\phi_\bk+(\sin2\theta_\bk\sin^{2}\phi_\bk)/2$\\
  \hline\hline
\end{tabular}
  \caption{ Form factors $\mathcal{F}^{\mu\nu}_{\alpha\beta}$. $\alpha$ and $\beta$ represent the different spin projections $(x,y,z)$, and $\mu,\nu$ indicate the band index, and take the values $\pm1$ [Eq.~\eqref{energies}]. Here we note that the $xz,yz$ and $yx$ form factors are the complex conjugates of the $zx,zy$ and $xy$ form factors respectively.}
\label{table}
\end{table}

The angular dependence of the spin susceptibility integrand [Eq.~\eqref{chispin2}] is limited to the dependence of the form factors on $\theta_\bk$. Hence, one can easily deduce from Table~\ref{table} that, upon tracing out the angular part, $\chi_{zx}(q_z)$$=$$\chi_{xz}(q_z)$$=$$\chi_{zy}(q_z)$$=$$\chi_{yz}(q_z)$$=$$0$. Additionally, we notice that {\it all} the diagonal elements of the spin susceptibility tensor $\chi_{ii}(q_z)$ $(i=x,y,z)$ have contributions from intraband and interband scattering processes. This is unlike the non-irradiated case, {\it i.e.} $\phi_\bk=\pi/2$ in Table~\ref{table}, where $\chi_{zz}(q_z)$ has contributions that are limited to interband transitions~\cite{Asmar2019}. Strikingly, the non-diagonal elements of the spin susceptibility, $\chi_{xy}(q_z)$ and $\chi_{yx}(q_z)$, are only displayed in the irradiated case. This can be seen by the angular integration in Eq.~\eqref{chispin2}, upon which the only non-zero term that contributes to these susceptibility elements is the interband term $\sim i\mu \cos\phi_\bk$, which vanishes in the non-irradiated case~\cite{Asmar2019}.

Our analysis reveals that BiTeI's spin susceptibility shows several qualitative differences under irradiation. These differences stem from the photo-generation of non-diagonal elements of the spin susceptibility, which in turn lead to the generation of non-collinear magnetic exchange in the magnetic multilayer:
\begin{eqnarray}\label{internonsim2}
{\rm J}(z)=-\frac{C}{2}&&\left[\sum_{\alpha, \beta=x,y}S^{\Top}_{\alpha}S^{\Bot}_{\beta}\int_{-\infty}^{\infty}{dq_{z}}e^{iq_{z}z}\chi_{\alpha\beta}(q_{z})\right.\nonumber \\
&&\left.+S^{\Top}_{z}S^{\Bot}_{z}\int_{-\infty}^{\infty}{dq_{z}}e^{iq_{z}z}\chi_{zz}(q_{z})\right]\;,
\end{eqnarray}
where $C=\left[J_{0}/(\mu_{\rm B}V_{0})\right]^2c^{2}/(2\pi)$.

In what follows we show that the interlayer exchange coupling of the system in Fig.~\ref{fig1}(a) will take the general form
\begin{eqnarray}\label{general}
{\rm J}(z)=\sum_{i=x,y,z}{\rm J}_{ii}(z)S^{\Top}_{i}S^{\Bot}_{i}+{\rm J}_{xy}(z)({\bm S}^{\Top}\times {\bm S}^{\Bot})\cdot{\hat{z}}\;.
\end{eqnarray}
Here ${\rm J}_{ii}(z)$ are the collinear magnetic exchange couplings, while ${\rm J}_{xy}$ is the light-induced non-collinear exchange coupling. We also find the $z$-dependence of the collinear and non-collinear exchange couplings and determine their periods of oscillation and decay power laws.

\subsection{Collinear Interlayer Magnetic Exchange}

\begin{figure}
  \centering
  \includegraphics[width=\columnwidth]{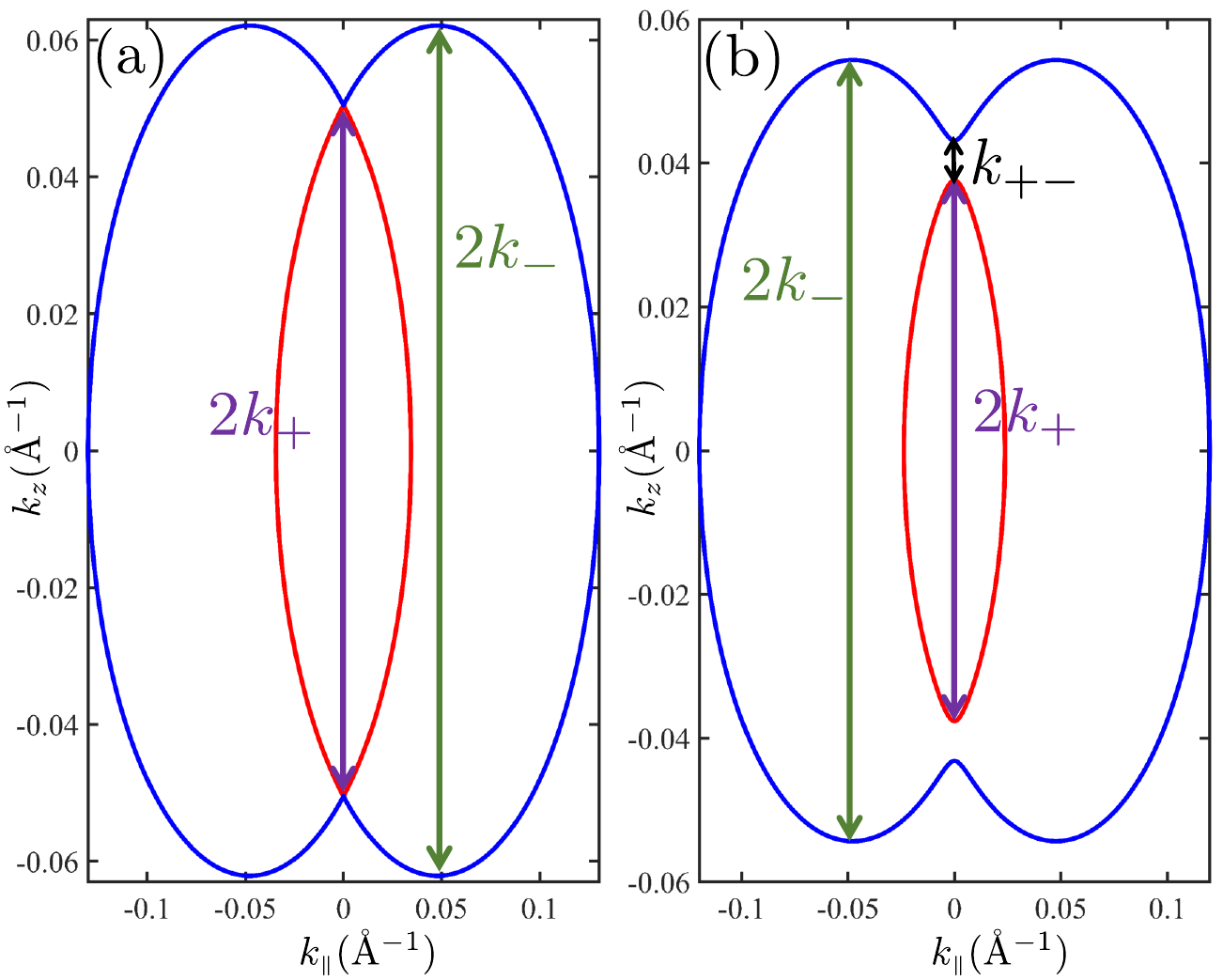}
  \caption{(a) Fermi surface projection of non-irradiated BiTeI displaying the nesting vectors $2k_{\pm}$. (b) Fermi surface of BiTeI under irradiation. The nesting vectors $2k_{\pm}$ in the irradiated case are shorter than those in the non-irradiated case due to the light-generated potential shift $\mathcal{V}$. A new nesting vector $k_{+-}$ appears due to the light-generated gap $2\Delta$.}\label{fig3}
\end{figure}

To start we consider the case where the spins of the ferromagnetic layers in Fig.~\ref{fig1}(a) are polarized in the $z$-direction. In this case, the magnetic exchange interaction between these two layers depends on the $\chi_{zz}(q_z)$ component of spin susceptibility, as indicated by  Eq.~\eqref{internonsim2}. At equilibrium, the helical states of BiTeI must undergo transitions that preserve the in-plane momentum $\bkpar$ in order to have non-zero contributions to the spin susceptibility [Eq.~\eqref{internonsim1}]. This constraint is reflected in $\chi_{zz}$ through the lack of intraband contributions at equilibrium (this can be seen in Table~\ref{table} by setting $\phi_{\bk}=\pi/2$), since the interaction of BiTeI's electrons with the $z$-polarized spins in the ferromagnets requires $\bkpar$-preserving spin flips. These spin flips can be only achieved by helicity flips that require the transition between bands~\cite{Asmar2019}. Moreover, the oscillatory nature of the RKKY-mediated interlayer exchange interaction results from the spin density oscillations induced by the ferromagnetic layers. These oscillations have the same physical origin as the Kohn anomaly~\cite{kohnanomaly}, which is induced due to the sharpness of the Fermi surface at zero temperature and is measured by the nesting vectors. In our system, nesting vectors connect two extremal points in the Fermi surface along the $k_z$-direction, as shown in Fig.~\ref{fig3}. Hence, from the shape of the Fermi surface at equilibrium, Fig.~\ref{fig3}(a), and since $\chi_{zz}(q_z)$ is limited to the contributions from interband transitions, one can deduce that the magnetic exchange coupling will have a period of oscillation determined by the nesting vector $2k_{+}$ in Fig.~\ref{fig3}(a), and given by $\pi/k_{+}$. Additionally, to obtain the dependence of ${\rm J}_{zz}$ on the separation of the magnetic layers $z$, we make use of the Riemann-Lebesgue lemma which states that if a function oscillates rapidly around zero then the integral of this function is small and the principal contribution to the integral arises from the behavior of the integrand in the vicinity of the non-analytic points. The non-analytic point in the integrand of the $zz$-component in Eq.~\eqref{internonsim2} is the nesting vector $2k_+$ (the Kohn anomaly). By performing the integral in the vicinity of this point we find that without irradiation ${\rm J}_{zz}(z)$ displays a behavior similar to that of a conventional 3D electron gas~\cite{LayersRKKY1,LayersRKKY2, Asmar2019}, since ${\rm J}_{zz}(z)\propto -(c/z)^{2} \cos(2k_{+}z)$ [$c$ is indicated in Fig.~\ref{fig1}(b)]. The details of the evaluation of the $\chi_{\alpha \beta}(q_z)$ components and the use of the Riemann-Lebesgue lemma to find the asymptotic form of the exchange are relegated to~\ref{Appendix1}.

When a BiTeI magnetic multilayer with $z$-polarized magnetic layers is subjected to a circularly polarized light the behavior of the interlayer magnetic exchange coupling is influenced through two main effects introduced by light. First, light-driving leads to a change in the shape of the Fermi surface, as it leads to the opening of a gap in its electronic spectrum, the rescaling of the Fermi level through a light-generated potential shift, and the renormalization of the Rashba SOC. The second effect of light is the modification of the low-energy electronic states' spin textures, which renders these states non-helical. These two light-generated effects lift the constraints imposed on the $\bkpar$-preserving scattering events (imposed by helicity at equilibrium) that contribute to the $\chi_{zz}(q_z)$ spin susceptibility component. This allows for new nesting vectors, and consequently leads to light-generated periods of oscillation that are also modulated by the light-induced potential shift. One can clearly notice the lifting of the helicity-imposed constraints on $\bkpar$-preserving scattering after the light is shone by setting $\phi_\bk\ne \pi/2$ in Table~\ref{table}. The non-helical electrons of irradiated BiTeI do not have conserved spin projections along the $k_z$-direction [Fig.~\ref{fig2}(c)] which allows the $\bkpar$-preserving transition within a single band or through two different ones to mediate the interaction between $z$-polarized ferromagnets. Therefore, under irradiation $\chi_{zz}(q_z)$ has contributions from both interband and intraband processes.
Moreover, noticing the Fermi surface shape of BiTeI after irradiation, Fig.~\ref{fig3}(b), and as both the interband and intraband transitions become non-vanishing in $\chi_{zz}(q)$, one can expect that in this case the interlayer exchange coupling will have three periods of oscillation determined by the three nesting vectors $2k_{\pm}$ and $k_{+-}$ shown in Fig.~\ref{fig3}(b).

The asymptotic form of ${\rm J}_{zz}(z)$ interlayer magnetic exchange coupling is found from $\chi_{zz}(q_z)$ in~\ref{Appendix1} and by using the Riemann-Lebesgue lemma,
\begin{figure}[t]
  \centering
  \includegraphics[width=\columnwidth]{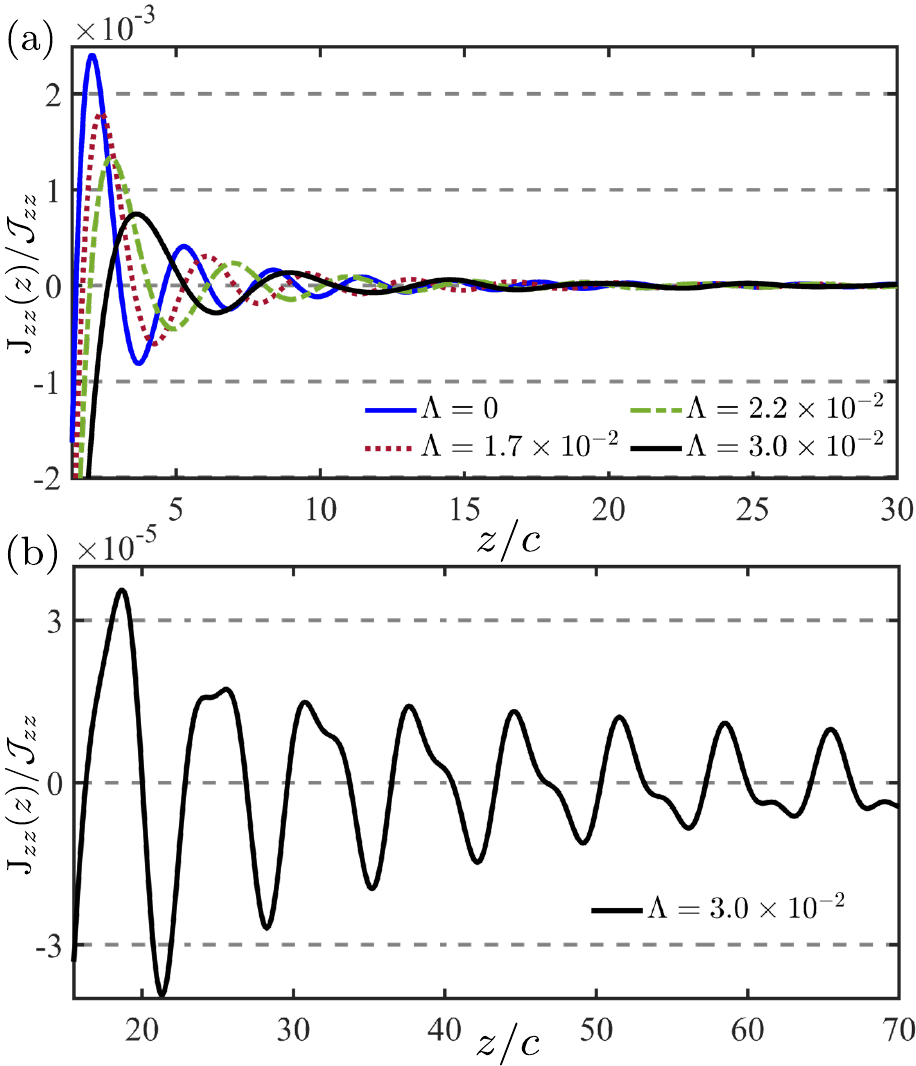}
  \caption{(a) Interlayer exchange coupling as a function of the separation between two ferromagnetic layers with magnetization normal to the $x-y$ plane of the system in Fig.~\ref{fig1}. The distinct plots in panel (a) correspond to different light-matter coupling strengths. (b) Closeup to the large $z$ region showing  the competition between the different periods of oscillation for the strongest light-matter coupling considered in (a). In both panels (a) and (b) the Fermi energy of the system is $E_{F}=180$ meV, and the incoming photon energy is $\hbar\Omega=7.5$ eV [$\mathcal{J}_{zz}$ is given in Eq.~\eqref{calJ}].}\label{fig4}
\end{figure}
\begin{eqnarray}\label{jzzlight}
&&{\rm J}_{zz}(z)\approx\mathcal{J}_{+1}\left(\frac{c}{z}\right)\cos(2k_+z)+\mathcal{J}_{+2}\left(\frac{c}{z}\right)^2\sin(2k_+z) \nonumber\\
&&+\mathcal{J}_{-}\left(\frac{c}{z}\right)\cos(2k_{-}z) +\mathcal{J}_{+-}\left(\frac{c}{z}\right)\cos(k_{+-}z),
\end{eqnarray}
where we have defined
\begin{subequations}\label{Js}
\begin{eqnarray}
&&\mathcal{J}_{+1}=-\frac{\mathcal{J}_{zz}\Delta^2 B c}{2A\alpha_{\rm R}^2}\left[\frac{1}{2k_{+}}-\left(\frac{2k_+\lambda_{-}}{\lambda_{+}(k_++k_{z-})}\right)\right]\;,\\
&&\mathcal{J}_{+2}=-\frac{\mathcal{J}_{zz}A}{4ABk_{+}+\alpha_{\rm R}^2+\Delta A}\left(2k_{+}+\frac{\Delta^2}{2\alpha_{\rm R}^2k_{+}}\right)\;,\\
&&\mathcal{J}_{-}=-\frac{\mathcal{J}_{zz}\Delta^2 B c}{4k_{-}A\alpha_{\rm R}^2},\;\mathcal{J}_{+-}=\frac{\mathcal{J}_{zz}\Delta^2 B c}{2A\alpha_{\rm R}^2}\left(\frac{k_{+-}\lambda_{-}}{\lambda_{+}(k_++k_{z-})}\right),\nonumber\\
\end{eqnarray}
\end{subequations}
$A$, $B$, $\alpha_{\rm R}$, and $\Delta$ are given in Eq.~\eqref{HeffSys}, and
\begin{equation}\label{calJ}
\mathcal{J}_{\alpha\beta}=\frac{S^{\Top}_{\alpha}S^{\Bot}_{\beta}}{16\pi B}\left(\frac{J_{0}}{V_{0}}\right)^2\;,
\end{equation}
$\lambda_{\pm}=\alpha_{\rm R}(1\pm\sqrt{4B(E_{F}-\mathcal{V})+\alpha^2_{\rm R}})/(2B)$. The wave vectors $k_{\pm}$ and $k_{+-}$ are
\begin{eqnarray}\label{wavevecs}
&&k_{+}=\sqrt{\frac{E_{F}-\mathcal{V}-\Delta}{A}}\;,\\
&&k_{-}=\frac{1}{2\alpha_{\rm R}}\sqrt{\frac{4B(E_{F}-\mathcal{V})\alpha_{\rm R}^2+\alpha_{\rm R}^4+(2B\Delta)^2}{AB}}\;,\nonumber
\end{eqnarray}
$k_{+-}=k_{z-}-k_{+}$, and $k_{z-}=\sqrt{(E_{F}-\mathcal{V}+\Delta)/A}$. Here we note that for consistency with the effective description of BiTeI we have restricted the values of $E_F<0.2$ eV (see Eq.~\eqref{jzzlight2} in~\ref{Appendix1} for the form of ${\rm J}_{zz}(z)$ for any $E_F$).

In Fig.~\ref{fig4}(a) we analyze the dependence of ${\rm J}_{zz}(z)$ on the light parameters through the variation of the dimensionless light-matter coupling $\Lambda$ [recall that, $\Lambda=\mathcal{A}\alpha_{\rm R}/(\hbar\Omega)$, $\mathcal{A}$ is given in Eq.~\eqref{HeffSys}].
Without irradiation the interlayer exchange coupling ${\rm J}_{zz}(z)$ has a single period of oscillation ($k_{+}=\sqrt{E_{F}/A}$) and it decays with the square of the separation of the magnetic layers, Fig~\ref{fig4}(a). The behaviour of ${\rm J}_{zz}(z)$ without light can be deduced from Eq.~\eqref{jzzlight} by setting $\Delta=\mathcal{V}=0$, $\alpha_{\rm R}={\alpha}_{\rm R}$, in Eqs.~\eqref{Js} and \eqref{wavevecs}. Irradiating the system leads to the generation of new oscillatory contributions in the $zz$-component of the interlayer exchange that become more prominent with increasing $\Lambda$, since they are proportional to the square of $\Delta=(\mathcal{A} \alpha_{\rm R})^2/(\hbar\Omega)$ as shown in Eq.~\eqref{Js}. These light-generated terms decay as the inverse of the separation between the magnetic layers, which resembles the coupling between two magnetic chains mediated by a 2DEG~\cite{kittelbook}. Due to the proportionality of the light-induced contributions to ${\alpha}_{\rm R}^2$ [Eq.~\eqref{Js}], we attribute the $z^{-1}$ dependence of these terms to the 2D nature of the Rashba SOC in BiTeI. Moreover, we notice that in ${\rm J}_{zz}(z)$ the period of oscillation resulting from the nesting vector $2k_{+}$ dominates, especially at small values of $z$, since this period is inherited from equilibrium. From Fig.~\ref{fig4} (a) it is clear that this period increases with $\Lambda$, since $k_{+}$ becomes smaller as $\mathcal{V}$ and $\Delta$ increase with increasing $\Lambda$ [Eq.~\eqref{wavevecs}]. The secondary periods associated with the nesting vectors $2k_{-}$ and $k_{+-}$ [Fig.~\ref{fig3} (b)], become more prominent for increasing light-matter couplings and large $z$-separations since their amplitudes increase as $\Delta^2$, and their decay envelope is $z^{-1}$. The effect of multiple competing oscillatory contributions with different periods and comparable amplitudes becomes evident in the interlayer exchange interaction for the $z$-range displayed in Fig.~\ref{fig4}(b).

\begin{figure}[t]
  \centering
  \includegraphics[width=\columnwidth]{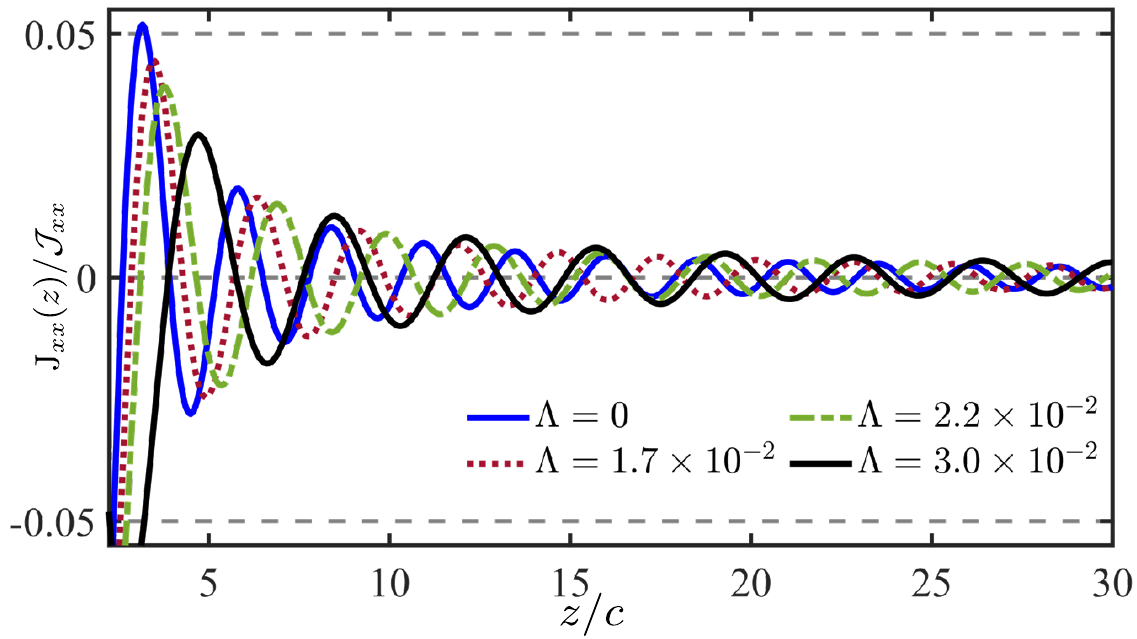}
  \caption{Dependence of the interlayer magnetic exchange interaction on the separation between magnetic layers with magnetization parallel to the $x-y$ plane of the system in Fig.~\ref{fig1}, for different light-matter coupling strengths. The Fermi energy and light frequency considered in this figure are given in Fig.~\ref{fig4}, we also note that ${\rm J}_{yy}(z)={\rm J}_{xx}(z)$ and $\mathcal{J}_{xx}$ is in Eq.~\eqref{calJ}.}\label{fig5}
\end{figure}
For a multilayer system composed of magnetic layers with in-plane magnetization, we find (\ref{Appendix1}) the asymptotic forms of ${\rm J}_{xx}(z)={\rm J}_{yy}(z)$, such that
\begin{eqnarray}\label{Jx}
{\rm J}_{xx}(z)\approx&&\frac{{\rm J}_{zz}(z)}{2}\\
&&+\mathcal{J}_{x,1}\left(\frac{c}{z}\right)\cos(2k_{-}z)+\mathcal{J}_{x,2}\left(\frac{c}{z}\right)^2\sin(2k_{-}z)\;.\nonumber
\end{eqnarray}
We have defined
\begin{equation}\label{Jsx}
  \mathcal{J}_{x,1} = -\frac{\mathcal{J}_{xx}(\alpha_{\rm R}^2+\Delta A)c}{8ABk_{-}},\;
  \mathcal{J}_{x,2}= -\frac{2\mathcal{J}_{xx}k_{-}\alpha_{\rm R}}{2(2k_{-}\alpha_{\rm R}+\Delta)}\;,
\end{equation}
and $\mathcal{J}_{xx}$ is given in Eq.~\eqref{calJ}. Here, we should note that even without irradiation the interlayer exchange interaction between ferromagnetic layers with polarizations parallel to the interface [${\rm J}_{xx}(z),{\rm J}_{yy}(z)$] result from the spin susceptibility components that have contributions from both the interband and the intraband transitions. Both of these contributions are present in $\chi_{xx}(q_{z})$ and $\chi_{yy}(q_{z})$ since the $\bkpar$-preserving transitions within or through the helical bands of non-irradiated BiTeI lead to the necessary spin flips required to mediate the interaction between the magnetic layers~\cite{Asmar2019}. Moreover, the asymptotic form of ${\rm J}_{xx/yy}(z)$ for the non-irradiated system (setting $\Delta=\mathcal{V}=0$ in Eqs.~\eqref{Jx} and \eqref{Jsx}) reveals that the interlayer exchange coupling oscillates with the periods associated with the nesting vectors $2k_{\pm}$ [Fig.~\ref{fig3} (a)] and has two decay powers, $z^{-2}$ and $z^{-1}$. Hence, the oscillations for this case are dominated by the term that is directly proportional to $\alpha_{\rm R}^2$, and ${\rm J}_{xx/yy}(z)\propto\alpha_{\rm R}^2(c/z)\sin(2k_{-}z)$, as seen in Fig.~\ref{fig5}. When we irradiate the system, the light-generated periods of oscillation will not have a significant effect even though they decay as $z^{-1}$, since their amplitudes are small compared to the amplitude inherited from equilibrium, i.e., the term proportional to $\mathcal{J}_{x,1}$ in Eq.~\eqref{Jx}. Then, as seen in Fig.~\ref{fig5}, the main effect of light on the interlayer magnetic exchange interactions ${\rm J}_{xx/yy}(z)$ is to increase the oscillation period $\pi/k_{-}$, since $k_{-}$  decreases with increasing $\Lambda$, as shown in Eq.~\eqref{wavevecs}.

In this part we have seen the effects of light on the collinear exchange interaction components, ${\rm J}_{zz}(z)$, ${\rm J}_{xx}(z)$, and ${\rm J}_{yy}(z)$. Through our analysis of the interlayer exchange interactions, we have shown that the effects of light-driving on the RKKY-mediated collinear (Ising) magnetic exchange coupling [${\rm J}_{zz}(z)$] is manifested by the appearance of new oscillation periods, decay powers, and an increase in the dominant oscillation wavelength for small separations. These new factors arise from the light-induced modification of the quasiparticles’ spin textures, the change of the Fermi surface shape, and the generation of a potential shift. On the other hand, we find that the effects of light on the (Heisenberg) magnetic exchange couplings ${\rm J}_{xx}(z)$ and ${\rm J}_{yy}(z)$, are limited to the modulation of a single dominant oscillation period, due to the light-generated gap and potential shift.

\subsection{Non-collinear Interlayer Magnetic Exchange}

When the magnetic layers are deposited across the stacking direction [$z$-direction in Fig.~\ref{fig1} (a)] of BiTeI, or when two magnetic impurities are embedded on a line perpendicular to the $x-y$ plane of BiTeI, the magnetic exchange interaction is limited to the collinear exchange couplings~\cite{Asmar2019,BiTeIsingleimps}. The absence of non-collinear exchange interactions is attributed to the helical nature of the quasiparticle excitations near the A-point which couples their in-plane momentum to their in-plane spins, Fig.~\ref{fig2} (a). Subjecting the BiTeI to circularly polarized light leads to the appearance of a light-generated gap. This gap is captured in the effective high-frequency Hamiltonian in Eq.~\eqref{HeffSys} through the Zeeman-like term $\Delta\sigma_z$. In addition to opening a gap in the electronic spectrum this light-induced term leads to the $z$-polarization of electron spins at $\bkpar=0$, i.e., the conduction and valence bands edges. As $\bkpar$ gradually increases, the out-of-plane component of the electron spins decreases while their helical in-plane components are gradually restored, as shown in Fig.~\ref{fig2} (c). The interplay between the spin and momentum directions across the Fermi surface leads to non-vanishing off-diagonal components in the spin susceptibility tensor ($\chi_{xy}(q_z)$ and $\chi_{yx}(q_z)$), as can be deduced from  Table~\ref{table} (after angular integration). Hence, light irradiation is responsible for the modification of the spin textures across the BiTeI Fermi surface, and the non-collinear (DM) magnetic exchange couplings are a property unique to the driven system in Fig.~\ref{fig1}(a).

We have obtained the asymptotic form of ${\rm J}_{xy}(z)$ and ${\rm J}_{yx}(z)$ through the determination of $\chi_{xy}(q_z)$ and $\chi_{yx}(q_z)$ in~\ref{Appendix1}, and the use of the Riemann-Lebesgue lemma for the non-analytic points, $2k_{+}$ and $k_{+-}$, of the integrand in Eq.~\eqref{internonsim2}. We find that ${\rm J}_{xy}(z)=-{\rm J}_{yx}(z)$,
\begin{eqnarray}\label{Jxy}
{\rm J}_{xy}(z)\approx\mathcal{J}_{xy,1}\left(\frac{c}{z}\right)\cos(2k_+z)+\mathcal{J}_{xy,2}\left(\frac{c}{z}\right)\cos(k_{+-}z),\nonumber\\
\end{eqnarray}
where
\begin{eqnarray}\label{Jsxy}
&&\mathcal{J}_{xy,1}=\mathcal{J}_{xy}\left(\frac{\Delta^2}{2Ak_{+}\alpha_{\rm R}^2}+ \frac{\Delta k_{+}}{4 A \alpha_{\rm R}^2(k_{+}+k_{z-})}\right)\nonumber,\\
&&\mathcal{J}_{xy,2}=-\mathcal{J}_{xy}\frac{\Delta k_{+-} c}{4 A \alpha_{\rm R}(k_{+}+k_{z-})}\;.
\end{eqnarray}

\begin{figure}[t]
  \centering
  \includegraphics[width=\columnwidth]{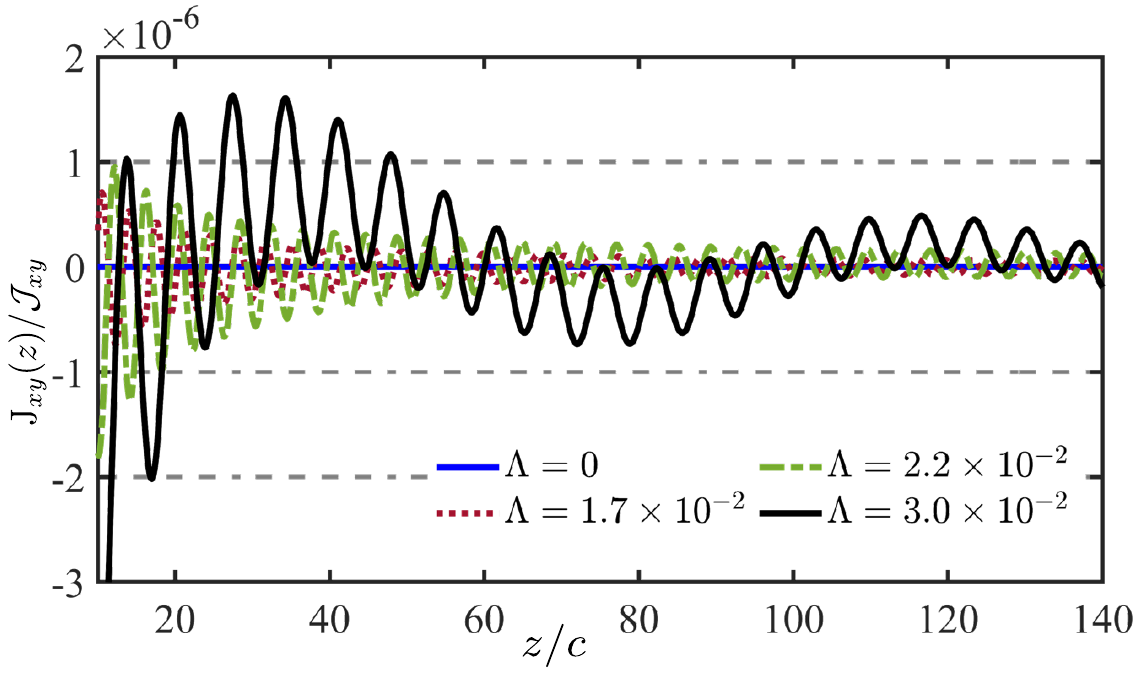}
  \caption{Non-collinear exchange interaction ${\rm J}_{xy}$ as a function of the BiTeI thickness $z$ for different values of light-matter coupling [note that ${\rm J}_{yx}(z)=-{\rm J}_{xy}(z)$]. The Fermi energy and the light frequency are given in Fig.~\ref{fig4}. }\label{fig6}
\end{figure}

In Fig.~\ref{fig6} we notice the vanishing of ${\rm J}_{xy}(z)$ and ${\rm J}_{yx}(z)$ without irradiation. With an increasing light-matter coupling, we can notice that the amplitude of the oscillations of the non-collinear terms increases, since these terms are proportional to $\Delta$, Eq.~\eqref{Jsxy}. From Eq.~\eqref{Jxy} we can notice that the non-collinear exchange has two periods of oscillations corresponding to the nesting vector $k_{+-}$ and $2k_{+}$. We also notice that the short period $\pi/k_{+}$ is dominant at small values of light-matter coupling considered in Fig.~\ref{fig6}, since $\mathcal{J}_{xy,2}$ is smaller than $\mathcal{J}_{xy,1}$ because $\Delta k_{+-}<\Delta^2$. The effect of the long period $2\pi/k_{+-}$ becomes more pronounced at large values of the light-matter coupling $\Lambda$. The latter is due to the increase in the oscillation amplitude to the point that it leads to small but visible modulations in ${\rm J}_{xy}(z)$ and ${\rm J}_{yx}(z)$, as shown in Fig.~\ref{fig6}.

In this section, we have seen that the irradiation of the BiTeI magnetic multilayer generates a DM magnetic exchange coupling which is, otherwise, absent without irradiation. The amplitude of the non-collinear exchange increases with increasing light-matter coupling and it becomes modulated by a secondary period at large values of this coupling.

\section{Conclusions}

We have presented our theoretical findings for light-driven effects on the magnetic exchange coupling between two magnetic layers deposited at opposite ends of the Rashba semiconductor BiTeI along the stacking direction. Our theory captures the light-induced modifications to BiTeI's Fermi surface and the electronic spin textures via an effective static Hamiltonian obtained from a high-frequency expansion of the non-perturbative Floquet Hamiltonian. Our work highlights the qualitative changes induced by light in the magnetic exchange coupling in Rashba magnetic multilayers, which make this platform promising for the exploration of periodic-driving effects on magnetic exchange interactions.

\begin{figure}[t]
  \centering
  \includegraphics[width=0.8\columnwidth]{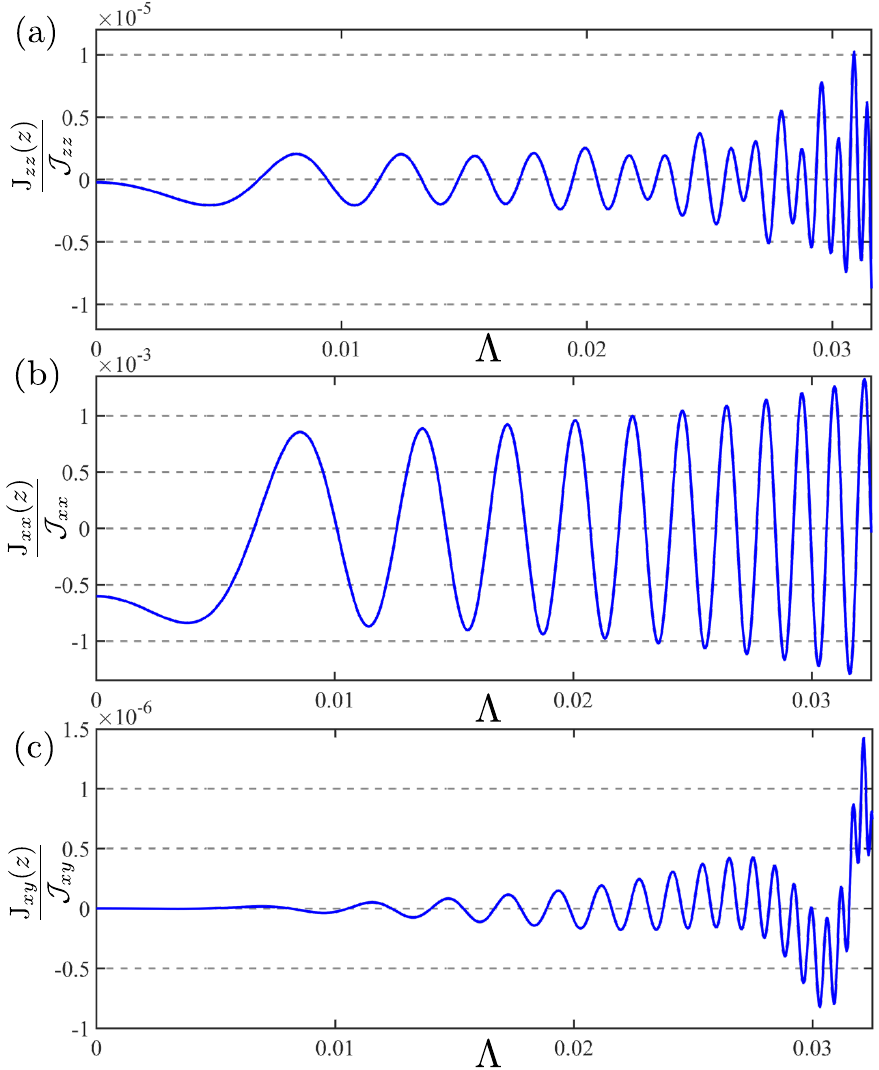}
  \caption{Light-driving contributions to the (a) Ising-like, (b) Heisenberg-like, and (c) DM-like interlayer exchange couplings as a function of the light-matter coupling for a fixed frequency, $\hbar\Omega=7.5$ eV and sample size $z=70$c. }\label{Newfig7}
\end{figure}

For magnetic layers with collinear magnetization, our calculations of the exchange interaction for a multilayer system under irradiation reveal that the Ising ($zz$) magnetic coupling acquires new periods of oscillations that arise from the light-induced change introduced to BiTei's Fermi surface shape, and the relaxation of the scattering constraints required by helicity for non-irradiated Rashba semiconductors. On the other hand, the main effect of irradiation on the Heisenberg ($xx,yy$) magnetic exchange coupling is the modulation of the oscillation periods due to the photon-generated gap and the potential shift. For both the Ising and Heisenberg magnetic exchange couplings their light-induced contributions decay with the inverse thickness of BiTeI.

Unlike a non-irradiated Rashba magnetic multilayer, with irradiation we find that there is a Dzyaloshinskii-Moriya exchange interaction that couples magnetic layers with non-collinear magnetization. This type of exchange arises from the light-induced variation of the electron spin orientations across the Fermi surface. We also find that the non-collinear exchange coupling increases with light-matter coupling, is characterized by two periods of oscillation, and decays as the inverse of the magnetic layer's separation.
 
Our predicted effects of light illumination on the interlayer exchange coupling can be measured via magneto-resistance oscillations~\cite{mangetores}, polarized neutron reflectrometry~\cite{neutorn}, ferromagnetic resonance~\cite{FerrRess}, and the magnetooptical Kerr effect~\cite{magopticalkerr}. Aside from using wedge-shaped samples~\cite{book-wedge} which require state-of-the-art fabrication techniques, here we propose an alternative strategy that does not involve changing the thickness of the BiTeI sample. The light-matter coupling depends on the light parameters as $\Lambda\propto E_{0}/(\Omega)$, where $E_{0}$ is the amplitude of the electric field component and $\Omega$ is the frequency of the irradiated light. Since we have considered the off-resonant, high-frequency regime, based on the skin depth of the Rashba semiconductor and the light frequency, we have determined the bounds on the sample thickness. Therefore, to achieve an experimental setup consistent with our theoretical considerations, the sample thickness and frequency are kept constant. Hence, the light intensity can be experimentally used to measure the RKKY oscillations and extract the decay envelopes without changing the light frequency or the sample
thickness. From Figs.~\ref{Newfig7} (a) and (c) we can notice the interlayer exchange coupling is dominated by a single period at relatively small values of $\Lambda$, and that the competition between two oscillatory terms with distinct periods becomes clear at large values of $\Lambda$. Moreover, in these
two figures we notice an increasing interlayer exchange coupling at large $\Lambda$, since for the fixed sample thickness $z$ the contribution from the light-generated terms that decay as $z^{-1}$ becomes dominant. In Fig.~\ref{Newfig7} (b)  we notice the dominance of a single period of oscillations and that there is no significant increase in the indirect exchange interaction at large values of $\Lambda$, since in this case the exchange is dominated by the $z^{-1}$ dependent terms resulting from the 2D Rashba SOC. The measurement scheme suggested in Fig.~\ref{Newfig7} is equivalent to simultaneously varying the sample size and the amplitudes of the oscillatory terms, since increasing $\Lambda$ tunes the arguments of the oscillatory functions and the light-matter coupling-dependent coefficients in Eqs.~\eqref{jzzlight},~\eqref{Jx}, and~\eqref{Jxy}. We also note that the continuous tuning of light polarization also leads to the variation of the nesting vectors and light-matter dependent coefficients since, as shown in Ref.~\cite{FloqTIReview}, the gap generated for a high-frequency light is maximum for circularly polarized light, is reduced for elliptical polarization and vanishes for linearly polarized light. Hence, for a fixed sample size, frequency and intensity of light, one can tune the light polarization to tune the wave vectors and amplitudes of the oscillatory terms of the interlayer exchange via the light-generated gap $\Delta$.

Our study presents a system which displays qualitative differences in its magnetic exchange couplings under irradiation. Specifically, the generation of non-collinear magnetic exchange couplings can serve as a clear experimental probe of periodic driving effects and Floquet engineering of quantum matter, since this kind of exchange is uniquely generated by irradiation in our proposed system. The theory and findings we
obtained for the BiTeI system are also applicable to other bulk Rashba semiconductors and spin-orbit coupled materials, and we expect that it will motivate future experimental and theoretical studies seeking the exploration of systems with phenomena unique to periodic-driving out of equilibrium.

$\;$

\section{Acknowledgements}
The authors acknowledge the insightful discussions with Takashi Mori and Thomas Iadecola.
This work was supported by the U.S. Department of Energy, Office of Science, Basic Energy Sciences under Early Career Award No. DE-SC0019326 (W-K.T.), and by the National Science Foundation via Grant No. DMR-2213429
(M.M.A.).

\section{Data Availability Statement}
 The data that support the findings of this study are available upon reasonable request from the authors.
\appendix

\begin{widetext}
\section{Spin susceptibility tensor components, asymptotic approximation, and comparison to numerical evaluations}\label{Appendix1}
In this section we outline the procedure followed in the derivation of the asymptotic forms of the interlayer magnetic exchange coupling components in the main text, Eqs.~\eqref{jzzlight},\eqref{Jx}, and \eqref{Jxy}. We start by finding the different non-zero susceptibility components, $\chi_{zz}(q_z)$, $\chi_{xx}(q_z)$, $\chi_{yy}(q_z)$, $\chi_{xy}(q_z)$ and $\chi_{yx}(q_z)$. For the $zz$-component we have,
\begin{eqnarray}\label{integralzz}
\chi_{zz}(q_z)= 2\mathcal{C}\int_{-\kappa }^{\kappa }dk_{z} \sum_{\mu}{\mathcal{P}\int_{0}^{k_{{ \scriptscriptstyle \parallel},\mu}}\left[\frac{1}{E_{\bk+q_{z},-\mu}-E_{\bk,\mu}}+\frac{1}{E_{\bk-q_{z},-\mu}-E_{\bk,\mu}}\right]\sin^2 \phi_\bk \kpar} d\kpar\nonumber\\
\;\;\;\;+\mathcal{C} \sum_{\mu} \int_{-k_{\mu} }^{k_{\mu} }dk_{z}{\mathcal{P}\int_{0}^{k_{{\scriptscriptstyle \parallel},\mu}}\left[\frac{1}{E_{\bk+q_{z},\mu}-E_{\bk,\mu}}+\frac{1}{E_{\bk-q_{z},\mu}-E_{\bk,\mu}}\right]\cos^2 \phi_\bk \kpar} d\kpar\;,
\end{eqnarray}
where $\kappa=k_{+}+k_{+-}$, $k_{\pm}$ and $k_{+-}$ are given in Eq.~\eqref{wavevecs}, $E_{\bk,\mu}$ is given in Eq.~\eqref{energies}, $\phi_{\bk}$ in Eq.~\eqref{basis}, $\mathcal{C}=\pi \mu^{2}_{\rm B}/(2\pi)^2$, and $\mathcal{P}$ denotes the principal value of the integral.

For $\chi_{xx}(q_{z})$ and $\chi_{yy}(q_{z})$, we notice from the dependence of their form factors on $\theta_\bk$ in Table~\ref{table}, that integrating over $\theta_\bk$ in Eq.~\eqref{spinsucesfeqfinal}, renders $\chi_{xx}(q_{z})=\chi_{yy}(q_{z})$, and
\begin{eqnarray}\label{integralxx}
\chi_{xx}(q_z)=\mathcal{C} \sum_{\mu} \int_{-k_{\mu} }^{k_{\mu} }dk_{z}{\mathcal{P}\int_{0}^{k_{{ \scriptscriptstyle \parallel},\mu}}\left[\frac{1}{E_{\bk+q_{z},\mu}-E_{\bk,\mu}}+\frac{1}{E_{\bk-q_{z},\mu}-E_{\bk,\mu}}\right]\sin^2 \phi_\bk \kpar} d\kpar \nonumber\\
\;\;\;+\mathcal{C}\int_{-\kappa }^{\kappa }dk_{z} \sum_{\mu}{\mathcal{P}\int_{0}^{k_{{ \scriptscriptstyle \parallel},\mu}}\left[\frac{1}{E_{\bk+q_{z},-\mu}-E_{\bk,\mu}}+\frac{1}{E_{\bk-q_{z},-\mu}-E_{\bk,\mu}}\right](\cos^2 \phi_\bk +1)\kpar} d\kpar.\nonumber\\
\end{eqnarray}

The allowed off-diagonal components of the spin susceptibility, after angular integration (see Table~\ref{table}), satisfy $\chi_{xy}(q_z)=-\chi_{yx}(q_z)$ and are given by
\begin{equation}\label{integralxy}
\chi_{xy}(q_z)=\mathcal{C}\int_{-\kappa }^{\kappa }dk_{z} \sum_{\mu}{\mathcal{P}\int_{0}^{k_{{ \scriptscriptstyle \parallel},\mu}}\left[\frac{1}{E_{\bk+q_{z},-\mu}-E_{\bk,\mu}}+\frac{1}{E_{\bk-q_{z},-\mu}-E_{\bk,\mu}}\right]i\mu \cos \phi_\bk \kpar} d\kpar\;.
\end{equation}

Performing the integrals in Eq.~\eqref{integralzz}$-$~\eqref{integralxy}, we get
\begin{eqnarray}\label{chizz}
\chi_{zz}(q_z)\approx  2\mathcal{C}\left\{\frac{\Delta^2}{2A\alpha^2_{\rm R}q_z}\left[\sum_{\mu}\ln\left|\frac{q_z+2k_{\mu}}{q_z-2k_{\mu}}\right| +\frac{\lambda_{+}q_z}{\lambda_{-}\left(k_{+}+k_{z,-}\right)}\ln\left|\frac{q_z+k_{z-}-k_{+}}{q_z-2k_+}\right| \right]+\frac{k_+}{B}\right.\nonumber\\
 \;\;\;\;\;\; \;\;\;\;\;\; \left.+\frac{A}{8}\left[q_z+\frac{\Delta^2}{2\alpha^2_{\rm R}q_z}\right]\left[\frac{4k^{2}_{+}+q^2_z}{\alpha^2_{\rm R}+A\Delta} +\frac{q^{2}_z-4k^2_+}{ABq^{2}_z+\alpha^2_{\rm R}+A\Delta}\right]\ln\left|\frac{2k_++q_z}{2k_-q_z}\right|\right.\nonumber\\
 \;\;\;\;\;\;\;\;\;\;\;\;  \left. +\left[ \frac{4k^{2}_{+}+q^2_z}{\alpha^2_{\rm R}+A\Delta}+\frac{4k^2_+-q^{2}_z}{ABq^{2}_z+A\Delta} \right]\ln\left|\frac{q_z(2k_+-q_z)-2(\alpha^2_{\rm R}+A\Delta)/(AB)}{q_z(2k_++q_z)+2(\alpha^2_{\rm R}+A\Delta)/(AB)}\right|\right. \nonumber\\ 
\left. +\frac{Ak_+q^2_z}{2(\alpha^2_{\rm R}+A\Delta)}\ln\left|\frac{q^2_z(q^2_z-4k^2_+)}{[q_z(2k_+-q_z)-2(\alpha^2_{\rm R}+A\Delta)/(AB)][q_z(2k_+-q_z)+2(\alpha^2_{\rm R}+A\Delta)/(AB)]}\right|\right\},\nonumber\\
\end{eqnarray}
\begin{eqnarray}\label{chixx}
 \chi_{xx}(q_z)\approx \frac{\chi_{zz}(q_z)}{2}+\frac{\mathcal{C}}{B}\left\{ k_-+\left[\frac{(4k^{2}_-q^2_z)}{4q_z+\Delta/\alpha_{\rm R}}+\frac{\alpha^2_{\rm R}+A\Delta}{4ABq_z} \right]\ln\left|\frac{q_z+2k_{-}}{q_z-2k_{-}}\right| \right\}\;,
\end{eqnarray}
and
\begin{equation}\label{chixy}
\chi_{xy}(q_z)\approx2\mathcal{C}\left[\frac{\Delta q_z}{4A \alpha_{\rm R}(k_++k_{z-})}\ln\left|\frac{q_z+k_{z-}-k_+}{q_z-2k_{+}}\right| -\frac{\Delta^2}{A\alpha^2_{\rm R}q_z}\ln\left|\frac{q_z+2k_+}{q_z-2k_{+}}\right| \right]\;.
\end{equation}
\end{widetext}
Here we note that after integrating over $\bkpar$, we approximated the result up to second order in $\Delta$ since this parameter is small when compared to other energy scales in the Hamiltonian due to the high-frequency approximation used to obtain the effective Hamiltonian in Eq.~\eqref{HeffSys}.
\begin{figure}
  \centering
  \includegraphics[width=\columnwidth]{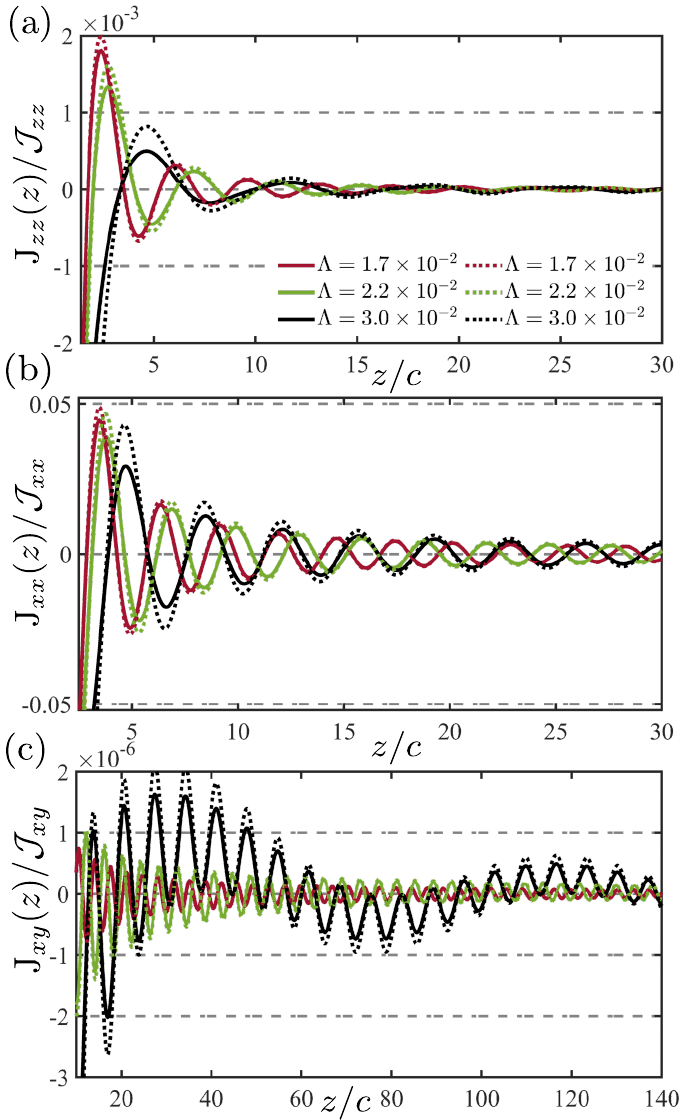}
  \caption{Comparison between the numerically evaluated exchange interactions (doted lines) and the approximate form in the main text (solid lines), Eqs.~\eqref{jzzlight}, ~\eqref{Jx}, and ~\eqref{Jxy}. ${\rm J}_{zz}(z)$, ${\rm J}_{xx}(z)$, and ${\rm J}_{xy}(z)$, are shown in panels (a), (b), and (c), respectively. The Fermi energy and light frequency used in our evaluations are given in Fig.~\ref{fig4}.}\label{fig7}
\end{figure}

Having found the spin susceptibility components we proceed to evaluate the interlayer exchange couplings in Eq.~\eqref{internonsim2}. From the expressions of the spin susceptibility components in Eqs.~\eqref{chizz}-\eqref{chixy}, one can notice that for certain values of $q_z$, either $\chi_{\alpha\beta}(q_z)$ or $\partial \chi_{\alpha\beta}(q_z)/\partial_{q_{z}}$ have logarithmic singularities
that are known as Kohn anomalies. The non-analytic behaviour of the spin susceptibility at these anomalous points, which are determined by the nesting vectors, allows us to use the Riemann-Lebesgue lemma. As an example of the use of the Riemann-Lebesgue lemma we consider the integral
\begin{equation}\label{example}
I(z)=\int_{-\infty}^{\infty} \cos(q_{z}z)\left(k_{+}-\frac{q^2_{z}-4k^2_{+}}{4q_{z}}\ln\left|\frac{q_z+2k_+}{q_z-2k_+}\right|\right)dq_z\;.
\end{equation}
In this integral the Kohn anomaly is displayed at $|q_z|=2k_+$, therefore the dominant contribution to the integral is obtained by integrating in the vicinity of the Kohn anomaly,
\begin{equation}\label{example2}
I(z)\approx \int_{2k_{+}-\epsilon}^{2k_{+}+\epsilon}\left(k_{+}+\frac{q_z-2k_+}{2}\ln\left|{q_z-2k_{+}}\right|\right)dq_z\;,
\end{equation}
where $\epsilon\ll 2k_{+}$. Changing variables to $q=q_z-2k_+$ and after integrating by parts we get
\begin{equation}\label{example3}
I(z)\approx \frac{\sin(2k_+z)}{z^2}\int_{-\epsilon}^{\epsilon}\frac{\sin(qz)}{q}dq\;,
\end{equation}
Moreover, as the main contribution of the previous integral comes from $q=0$, we can extend the integral to $\pm \infty$, and we get
\begin{equation}\label{example4}
I(z)\approx \frac{\pi \sin(2k_+z)}{ z^2}\;.
\end{equation}
The procedure outlined by the example provided above is followed to determine the asymptotic form on the interlayer exchange interaction in Eqs.~\eqref{jzzlight}, \eqref{Jx}, and ~\eqref{Jxy} in the main text. Here we should point out that without restrictions on the value of $E_F$, the interlayer exchange coupling ${\rm J}_{zz}(z)$, takes the form
\begin{eqnarray}\label{jzzlight2}
&&{\rm J}_{zz}(z)\approx \mathcal{J}_{+1}\left(\frac{c}{z}\right)\cos(2k_+z)+\mathcal{J}_{+2}\left(\frac{c}{z}\right)^2\sin(2k_+z)\nonumber\\
&&+\mathcal{J}_{-}\left(\frac{c}{z}\right)\cos(2k_{-}z) +\mathcal{J}_{+-}\left(\frac{c}{z}\right)\cos(k_{+-}z)\\
&&+\Theta\left(E_F-\frac{2\alpha^2_{\rm R}}{B}-\Delta\right)\sum_{\tau=\pm}\frac{k^{\tau}_{+}-k_{+}}{B(2k_{+}+k^{\tau}_{+})}\left(\frac{c}{z}\right)^2\sin(k^{\tau}_{+}z),\nonumber
\end{eqnarray}
where $k^{\tau}_{+}=k_{+}+\tau\sqrt{k^{2}_{+}-2[\alpha^2_{\rm R}/(AB)+\Delta/B]}$. However, sampling the oscillations associated with these nesting vectors requires $E_{F}>\frac{2\alpha^2_{\rm R}}{B}+\Delta$, i.e., $E_F\geq 0.7$ eV which is larger than the cutoff of the effective description of BiTeI.

In Fig.~\ref{fig7}, we compare the validity of our analytical approximate forms by comparing them to the direct numerical integration of Eqs.~\eqref{chispin2} and \eqref{internonsim2}. One can notice the good agreement between the ``exact'' numerical evaluations and our approximate forms for the interlayer exchange. The deviation of our approximation from the exact exchange interaction is more noticeable with increasing light-matter coupling. However, this deviation remains small as the validity of our theory requires high frequencies, which constraints $\Lambda$ to small values.


\section*{References}
\bibliography{References2}

\begin{thebibliography}{81}%
\makeatletter
\providecommand \@ifxundefined [1]{%
 \@ifx{#1\undefined}
}%
\providecommand \@ifnum [1]{%
 \ifnum #1\expandafter \@firstoftwo
 \else \expandafter \@secondoftwo
 \fi
}%
\providecommand \@ifx [1]{%
 \ifx #1\expandafter \@firstoftwo
 \else \expandafter \@secondoftwo
 \fi
}%
\providecommand \natexlab [1]{#1}%
\providecommand \enquote  [1]{``#1''}%
\providecommand \bibnamefont  [1]{#1}%
\providecommand \bibfnamefont [1]{#1}%
\providecommand \citenamefont [1]{#1}%
\providecommand \href@noop [0]{\@secondoftwo}%
\providecommand \href [0]{\begingroup \@sanitize@url \@href}%
\providecommand \@href[1]{\@@startlink{#1}\@@href}%
\providecommand \@@href[1]{\endgroup#1\@@endlink}%
\providecommand \@sanitize@url [0]{\catcode `\\12\catcode `\$12\catcode
  `\&12\catcode `\#12\catcode `\^12\catcode `\_12\catcode `\%12\relax}%
\providecommand \@@startlink[1]{}%
\providecommand \@@endlink[0]{}%
\providecommand \url  [0]{\begingroup\@sanitize@url \@url }%
\providecommand \@url [1]{\endgroup\@href {#1}{\urlprefix }}%
\providecommand \urlprefix  [0]{URL }%
\providecommand \Eprint [0]{\href }%
\providecommand \doibase [0]{https://doi.org/}%
\providecommand \selectlanguage [0]{\@gobble}%
\providecommand \bibinfo  [0]{\@secondoftwo}%
\providecommand \bibfield  [0]{\@secondoftwo}%
\providecommand \translation [1]{[#1]}%
\providecommand \BibitemOpen [0]{}%
\providecommand \bibitemStop [0]{}%
\providecommand \bibitemNoStop [0]{.\EOS\space}%
\providecommand \EOS [0]{\spacefactor3000\relax}%
\providecommand \BibitemShut  [1]{\csname bibitem#1\endcsname}%
\let\auto@bib@innerbib\@empty
\bibitem [{\citenamefont {Lloyd-Hughes}\ \emph {et~al.}(2021)\citenamefont
  {Lloyd-Hughes}, \citenamefont {Oppeneer}, \citenamefont {dos Santos},
  \citenamefont {Schleife}, \citenamefont {Meng}, \citenamefont {Sentef},
  \citenamefont {Ruggenthaler}, \citenamefont {Rubio}, \citenamefont {Radu},
  \citenamefont {Murnane}, \citenamefont {Shi}, \citenamefont {Kapteyn},
  \citenamefont {Stadtmüller}, \citenamefont {Dani}, \citenamefont
  {da~Jornada}, \citenamefont {Prinz}, \citenamefont {Aeschlimann},
  \citenamefont {Milot}, \citenamefont {Burdanova}, \citenamefont {Boland},
  \citenamefont {Cocker},\ and\ \citenamefont {Hegmann}}]{light0}%
  \BibitemOpen
  \bibfield  {author} {\bibinfo {author} {\bibfnamefont {J.}~\bibnamefont
  {Lloyd-Hughes}}, \bibinfo {author} {\bibfnamefont {P.~M.}\ \bibnamefont
  {Oppeneer}}, \bibinfo {author} {\bibfnamefont {T.~P.}\ \bibnamefont {dos
  Santos}}, \bibinfo {author} {\bibfnamefont {A.}~\bibnamefont {Schleife}},
  \bibinfo {author} {\bibfnamefont {S.}~\bibnamefont {Meng}}, \bibinfo {author}
  {\bibfnamefont {M.~A.}\ \bibnamefont {Sentef}}, \bibinfo {author}
  {\bibfnamefont {M.}~\bibnamefont {Ruggenthaler}}, \bibinfo {author}
  {\bibfnamefont {A.}~\bibnamefont {Rubio}}, \bibinfo {author} {\bibfnamefont
  {I.}~\bibnamefont {Radu}}, \bibinfo {author} {\bibfnamefont {M.}~\bibnamefont
  {Murnane}}, \bibinfo {author} {\bibfnamefont {X.}~\bibnamefont {Shi}},
  \bibinfo {author} {\bibfnamefont {H.}~\bibnamefont {Kapteyn}}, \bibinfo
  {author} {\bibfnamefont {B.}~\bibnamefont {Stadtmüller}}, \bibinfo {author}
  {\bibfnamefont {K.~M.}\ \bibnamefont {Dani}}, \bibinfo {author}
  {\bibfnamefont {F.~H.}\ \bibnamefont {da~Jornada}}, \bibinfo {author}
  {\bibfnamefont {E.}~\bibnamefont {Prinz}}, \bibinfo {author} {\bibfnamefont
  {M.}~\bibnamefont {Aeschlimann}}, \bibinfo {author} {\bibfnamefont {R.~L.}\
  \bibnamefont {Milot}}, \bibinfo {author} {\bibfnamefont {M.}~\bibnamefont
  {Burdanova}}, \bibinfo {author} {\bibfnamefont {J.}~\bibnamefont {Boland}},
  \bibinfo {author} {\bibfnamefont {T.}~\bibnamefont {Cocker}},\ and\ \bibinfo
  {author} {\bibfnamefont {F.}~\bibnamefont {Hegmann}},\ }\href
  {https://doi.org/10.1088/1361-648x/abfe21} {\bibfield  {journal} {\bibinfo
  {journal} {Journal of Physics: Condensed Matter}\ }\textbf {\bibinfo {volume}
  {33}},\ \bibinfo {pages} {353001} (\bibinfo {year} {2021})}\BibitemShut
  {NoStop}%
\bibitem [{\citenamefont {Mankowsky}\ \emph {et~al.}(2016)\citenamefont
  {Mankowsky}, \citenamefont {Först},\ and\ \citenamefont
  {Cavalleri}}]{light2}%
  \BibitemOpen
  \bibfield  {author} {\bibinfo {author} {\bibfnamefont {R.}~\bibnamefont
  {Mankowsky}}, \bibinfo {author} {\bibfnamefont {M.}~\bibnamefont {Först}},\
  and\ \bibinfo {author} {\bibfnamefont {A.}~\bibnamefont {Cavalleri}},\ }\href
  {https://doi.org/10.1088/0034-4885/79/6/064503} {\bibfield  {journal}
  {\bibinfo  {journal} {Reports on Progress in Physics}\ }\textbf {\bibinfo
  {volume} {79}},\ \bibinfo {pages} {064503} (\bibinfo {year}
  {2016})}\BibitemShut {NoStop}%
\bibitem [{\citenamefont {Oka}\ and\ \citenamefont
  {Kitamura}(2019)}]{flreview1}%
  \BibitemOpen
  \bibfield  {author} {\bibinfo {author} {\bibfnamefont {T.}~\bibnamefont
  {Oka}}\ and\ \bibinfo {author} {\bibfnamefont {S.}~\bibnamefont {Kitamura}},\
  }\href {https://doi.org/10.1146/annurev-conmatphys-031218-013423} {\bibfield
  {journal} {\bibinfo  {journal} {Annual Review of Condensed Matter Physics}\
  }\textbf {\bibinfo {volume} {10}},\ \bibinfo {pages} {387} (\bibinfo {year}
  {2019})}\BibitemShut {NoStop}%
\bibitem [{\citenamefont {Rudner}\ and\ \citenamefont
  {Lindner}(2020)}]{flreview2}%
  \BibitemOpen
  \bibfield  {author} {\bibinfo {author} {\bibfnamefont {M.~S.}\ \bibnamefont
  {Rudner}}\ and\ \bibinfo {author} {\bibfnamefont {N.~H.}\ \bibnamefont
  {Lindner}},\ }\href {https://www.nature.com/articles/s42254-020-0170-z}
  {\bibfield  {journal} {\bibinfo  {journal} {Nature Reviews Physics}\ }\textbf
  {\bibinfo {volume} {2}},\ \bibinfo {pages} {229} (\bibinfo {year}
  {2020})}\BibitemShut {NoStop}%
\bibitem [{\citenamefont {Basov}\ \emph {et~al.}(2017)\citenamefont {Basov},
  \citenamefont {Averitt},\ and\ \citenamefont {Hsieh}}]{flreview3}%
  \BibitemOpen
  \bibfield  {author} {\bibinfo {author} {\bibfnamefont {D.}~\bibnamefont
  {Basov}}, \bibinfo {author} {\bibfnamefont {R.}~\bibnamefont {Averitt}},\
  and\ \bibinfo {author} {\bibfnamefont {D.}~\bibnamefont {Hsieh}},\ }\href
  {https://doi.org/10.1038/nmat5017} {\bibfield  {journal} {\bibinfo  {journal}
  {Nature Materials}\ }\textbf {\bibinfo {volume} {16}},\ \bibinfo {pages}
  {1077} (\bibinfo {year} {2017})}\BibitemShut {NoStop}%
\bibitem [{\citenamefont {Aoki}\ \emph {et~al.}(2014)\citenamefont {Aoki},
  \citenamefont {Tsuji}, \citenamefont {Eckstein}, \citenamefont {Kollar},
  \citenamefont {Oka},\ and\ \citenamefont {Werner}}]{Oka_RMP}%
  \BibitemOpen
  \bibfield  {author} {\bibinfo {author} {\bibfnamefont {H.}~\bibnamefont
  {Aoki}}, \bibinfo {author} {\bibfnamefont {N.}~\bibnamefont {Tsuji}},
  \bibinfo {author} {\bibfnamefont {M.}~\bibnamefont {Eckstein}}, \bibinfo
  {author} {\bibfnamefont {M.}~\bibnamefont {Kollar}}, \bibinfo {author}
  {\bibfnamefont {T.}~\bibnamefont {Oka}},\ and\ \bibinfo {author}
  {\bibfnamefont {P.}~\bibnamefont {Werner}},\ }\href
  {https://doi.org/10.1103/RevModPhys.86.779} {\bibfield  {journal} {\bibinfo
  {journal} {Rev. Mod. Phys.}\ }\textbf {\bibinfo {volume} {86}},\ \bibinfo
  {pages} {779} (\bibinfo {year} {2014})}\BibitemShut {NoStop}%
\bibitem [{\citenamefont {Platero}\ and\ \citenamefont
  {Aguado}(2004)}]{Platero}%
  \BibitemOpen
  \bibfield  {author} {\bibinfo {author} {\bibfnamefont {G.}~\bibnamefont
  {Platero}}\ and\ \bibinfo {author} {\bibfnamefont {R.}~\bibnamefont
  {Aguado}},\ }\href
  {https://doi.org/https://doi.org/10.1016/j.physrep.2004.01.004} {\bibfield
  {journal} {\bibinfo  {journal} {Phys. Rep.}\ }\textbf {\bibinfo {volume}
  {395}},\ \bibinfo {pages} {1 } (\bibinfo {year} {2004})}\BibitemShut
  {NoStop}%
\bibitem [{\citenamefont {Oka}\ and\ \citenamefont {Aoki}(2009)}]{ftrans1}%
  \BibitemOpen
  \bibfield  {author} {\bibinfo {author} {\bibfnamefont {T.}~\bibnamefont
  {Oka}}\ and\ \bibinfo {author} {\bibfnamefont {H.}~\bibnamefont {Aoki}},\
  }\href {https://doi.org/10.1103/PhysRevB.79.081406} {\bibfield  {journal}
  {\bibinfo  {journal} {Phys. Rev. B}\ }\textbf {\bibinfo {volume} {79}},\
  \bibinfo {pages} {081406} (\bibinfo {year} {2009})}\BibitemShut {NoStop}%
\bibitem [{\citenamefont {Lee}\ and\ \citenamefont {Tse}(2017)}]{Floqtrans3}%
  \BibitemOpen
  \bibfield  {author} {\bibinfo {author} {\bibfnamefont {W.-R.}\ \bibnamefont
  {Lee}}\ and\ \bibinfo {author} {\bibfnamefont {W.-K.}\ \bibnamefont {Tse}},\
  }\href {https://doi.org/10.1103/PhysRevB.95.201411} {\bibfield  {journal}
  {\bibinfo  {journal} {Phys. Rev. B}\ }\textbf {\bibinfo {volume} {95}},\
  \bibinfo {pages} {201411} (\bibinfo {year} {2017})}\BibitemShut {NoStop}%
\bibitem [{\citenamefont {Rodriguez-Vega}\ \emph {et~al.}(2020)\citenamefont
  {Rodriguez-Vega}, \citenamefont {Vogl},\ and\ \citenamefont
  {Fiete}}]{martin1}%
  \BibitemOpen
  \bibfield  {author} {\bibinfo {author} {\bibfnamefont {M.}~\bibnamefont
  {Rodriguez-Vega}}, \bibinfo {author} {\bibfnamefont {M.}~\bibnamefont
  {Vogl}},\ and\ \bibinfo {author} {\bibfnamefont {G.~A.}\ \bibnamefont
  {Fiete}},\ }\href {https://doi.org/10.1103/PhysRevResearch.2.033494}
  {\bibfield  {journal} {\bibinfo  {journal} {Phys. Rev. Research}\ }\textbf
  {\bibinfo {volume} {2}},\ \bibinfo {pages} {033494} (\bibinfo {year}
  {2020})}\BibitemShut {NoStop}%
\bibitem [{\citenamefont {Asmar}\ and\ \citenamefont {Tse}(2022)}]{Asmar2022}%
  \BibitemOpen
  \bibfield  {author} {\bibinfo {author} {\bibfnamefont {M.~M.}\ \bibnamefont
  {Asmar}}\ and\ \bibinfo {author} {\bibfnamefont {W.-K.}\ \bibnamefont
  {Tse}},\ }\href {https://doi.org/10.1088/1361-648x/ac709d} {\bibfield
  {journal} {\bibinfo  {journal} {Journal of Physics: Condensed Matter}\
  }\textbf {\bibinfo {volume} {34}},\ \bibinfo {pages} {315602} (\bibinfo
  {year} {2022})}\BibitemShut {NoStop}%
\bibitem [{\citenamefont {Cayssol}\ \emph {et~al.}(2013)\citenamefont
  {Cayssol}, \citenamefont {D{\'o}ra}, \citenamefont {Simon},\ and\
  \citenamefont {Moessner}}]{FloqTIReview}%
  \BibitemOpen
  \bibfield  {author} {\bibinfo {author} {\bibfnamefont {J.}~\bibnamefont
  {Cayssol}}, \bibinfo {author} {\bibfnamefont {B.}~\bibnamefont {D{\'o}ra}},
  \bibinfo {author} {\bibfnamefont {F.}~\bibnamefont {Simon}},\ and\ \bibinfo
  {author} {\bibfnamefont {R.}~\bibnamefont {Moessner}},\ }\href
  {https://doi.org/10.1002/pssr.201206451} {\bibfield  {journal} {\bibinfo
  {journal} {Physica Status Solidi (RRL)--Rapid Research Letters}\ }\textbf
  {\bibinfo {volume} {7}},\ \bibinfo {pages} {101} (\bibinfo {year}
  {2013})}\BibitemShut {NoStop}%
\bibitem [{\citenamefont {Rudner}\ \emph {et~al.}(2013)\citenamefont {Rudner},
  \citenamefont {Lindner}, \citenamefont {Berg},\ and\ \citenamefont
  {Levin}}]{Floqtop3}%
  \BibitemOpen
  \bibfield  {author} {\bibinfo {author} {\bibfnamefont {M.~S.}\ \bibnamefont
  {Rudner}}, \bibinfo {author} {\bibfnamefont {N.~H.}\ \bibnamefont {Lindner}},
  \bibinfo {author} {\bibfnamefont {E.}~\bibnamefont {Berg}},\ and\ \bibinfo
  {author} {\bibfnamefont {M.}~\bibnamefont {Levin}},\ }\href
  {https://doi.org/10.1103/PhysRevX.3.031005} {\bibfield  {journal} {\bibinfo
  {journal} {Phys. Rev. X}\ }\textbf {\bibinfo {volume} {3}},\ \bibinfo {pages}
  {031005} (\bibinfo {year} {2013})}\BibitemShut {NoStop}%
\bibitem [{\citenamefont {Lindner}\ \emph {et~al.}(2011)\citenamefont
  {Lindner}, \citenamefont {Refael},\ and\ \citenamefont
  {Galitski}}]{Floqtop2}%
  \BibitemOpen
  \bibfield  {author} {\bibinfo {author} {\bibfnamefont {N.~H.}\ \bibnamefont
  {Lindner}}, \bibinfo {author} {\bibfnamefont {G.}~\bibnamefont {Refael}},\
  and\ \bibinfo {author} {\bibfnamefont {V.}~\bibnamefont {Galitski}},\ }\href
  {https://www.nature.com/articles/nphys1926} {\bibfield  {journal} {\bibinfo
  {journal} {Nat. Phys.}\ }\textbf {\bibinfo {volume} {7}},\ \bibinfo {pages}
  {490} (\bibinfo {year} {2011})}\BibitemShut {NoStop}%
\bibitem [{\citenamefont {de~la Torre}\ \emph {et~al.}(2021)\citenamefont
  {de~la Torre}, \citenamefont {Kennes}, \citenamefont {Claassen},
  \citenamefont {Gerber}, \citenamefont {McIver},\ and\ \citenamefont
  {Sentef}}]{sentef1}%
  \BibitemOpen
  \bibfield  {author} {\bibinfo {author} {\bibfnamefont {A.}~\bibnamefont
  {de~la Torre}}, \bibinfo {author} {\bibfnamefont {D.~M.}\ \bibnamefont
  {Kennes}}, \bibinfo {author} {\bibfnamefont {M.}~\bibnamefont {Claassen}},
  \bibinfo {author} {\bibfnamefont {S.}~\bibnamefont {Gerber}}, \bibinfo
  {author} {\bibfnamefont {J.~W.}\ \bibnamefont {McIver}},\ and\ \bibinfo
  {author} {\bibfnamefont {M.~A.}\ \bibnamefont {Sentef}},\ }\href
  {https://doi.org/10.1103/RevModPhys.93.041002} {\bibfield  {journal}
  {\bibinfo  {journal} {Rev. Mod. Phys.}\ }\textbf {\bibinfo {volume} {93}},\
  \bibinfo {pages} {041002} (\bibinfo {year} {2021})}\BibitemShut {NoStop}%
\bibitem [{\citenamefont {Dehghani}\ and\ \citenamefont
  {Mitra}(2015{\natexlab{a}})}]{opt2}%
  \BibitemOpen
  \bibfield  {author} {\bibinfo {author} {\bibfnamefont {H.}~\bibnamefont
  {Dehghani}}\ and\ \bibinfo {author} {\bibfnamefont {A.}~\bibnamefont
  {Mitra}},\ }\href {https://doi.org/10.1103/PhysRevB.92.165111} {\bibfield
  {journal} {\bibinfo  {journal} {Phys. Rev. B}\ }\textbf {\bibinfo {volume}
  {92}},\ \bibinfo {pages} {165111} (\bibinfo {year}
  {2015}{\natexlab{a}})}\BibitemShut {NoStop}%
\bibitem [{\citenamefont {Chen}\ \emph {et~al.}(2018)\citenamefont {Chen},
  \citenamefont {Du},\ and\ \citenamefont {Fiete}}]{opt1}%
  \BibitemOpen
  \bibfield  {author} {\bibinfo {author} {\bibfnamefont {Q.}~\bibnamefont
  {Chen}}, \bibinfo {author} {\bibfnamefont {L.}~\bibnamefont {Du}},\ and\
  \bibinfo {author} {\bibfnamefont {G.~A.}\ \bibnamefont {Fiete}},\ }\href
  {https://doi.org/10.1103/PhysRevB.97.035422} {\bibfield  {journal} {\bibinfo
  {journal} {Phys. Rev. B}\ }\textbf {\bibinfo {volume} {97}},\ \bibinfo
  {pages} {035422} (\bibinfo {year} {2018})}\BibitemShut {NoStop}%
\bibitem [{\citenamefont {Usaj}\ \emph {et~al.}(2014)\citenamefont {Usaj},
  \citenamefont {Perez-Piskunow}, \citenamefont {Torres},\ and\ \citenamefont
  {Balseiro}}]{Torres_graphene}%
  \BibitemOpen
  \bibfield  {author} {\bibinfo {author} {\bibfnamefont {G.}~\bibnamefont
  {Usaj}}, \bibinfo {author} {\bibfnamefont {P.~M.}\ \bibnamefont
  {Perez-Piskunow}}, \bibinfo {author} {\bibfnamefont {L.~F.}\ \bibnamefont
  {Torres}},\ and\ \bibinfo {author} {\bibfnamefont {C.~A.}\ \bibnamefont
  {Balseiro}},\ }\href {https://link.aps.org/doi/10.1103/PhysRevB.90.115423}
  {\bibfield  {journal} {\bibinfo  {journal} {Physical Review B}\ }\textbf
  {\bibinfo {volume} {90}},\ \bibinfo {pages} {115423} (\bibinfo {year}
  {2014})}\BibitemShut {NoStop}%
\bibitem [{\citenamefont {Topp}\ \emph {et~al.}(2019)\citenamefont {Topp},
  \citenamefont {Jotzu}, \citenamefont {McIver}, \citenamefont {Xian},
  \citenamefont {Rubio},\ and\ \citenamefont {Sentef}}]{sentef2}%
  \BibitemOpen
  \bibfield  {author} {\bibinfo {author} {\bibfnamefont {G.~E.}\ \bibnamefont
  {Topp}}, \bibinfo {author} {\bibfnamefont {G.}~\bibnamefont {Jotzu}},
  \bibinfo {author} {\bibfnamefont {J.~W.}\ \bibnamefont {McIver}}, \bibinfo
  {author} {\bibfnamefont {L.}~\bibnamefont {Xian}}, \bibinfo {author}
  {\bibfnamefont {A.}~\bibnamefont {Rubio}},\ and\ \bibinfo {author}
  {\bibfnamefont {M.~A.}\ \bibnamefont {Sentef}},\ }\href
  {https://doi.org/10.1103/PhysRevResearch.1.023031} {\bibfield  {journal}
  {\bibinfo  {journal} {Phys. Rev. Res.}\ }\textbf {\bibinfo {volume} {1}},\
  \bibinfo {pages} {023031} (\bibinfo {year} {2019})}\BibitemShut {NoStop}%
\bibitem [{\citenamefont {Kumar}\ \emph
  {et~al.}(2020{\natexlab{a}})\citenamefont {Kumar}, \citenamefont
  {Rodriguez-Vega}, \citenamefont {Pereg-Barnea},\ and\ \citenamefont
  {Seradjeh}}]{martin3}%
  \BibitemOpen
  \bibfield  {author} {\bibinfo {author} {\bibfnamefont {A.}~\bibnamefont
  {Kumar}}, \bibinfo {author} {\bibfnamefont {M.}~\bibnamefont
  {Rodriguez-Vega}}, \bibinfo {author} {\bibfnamefont {T.}~\bibnamefont
  {Pereg-Barnea}},\ and\ \bibinfo {author} {\bibfnamefont {B.}~\bibnamefont
  {Seradjeh}},\ }\href {https://doi.org/10.1103/PhysRevB.101.174314} {\bibfield
   {journal} {\bibinfo  {journal} {Phys. Rev. B}\ }\textbf {\bibinfo {volume}
  {101}},\ \bibinfo {pages} {174314} (\bibinfo {year}
  {2020}{\natexlab{a}})}\BibitemShut {NoStop}%
\bibitem [{\citenamefont {Trevisan}\ \emph {et~al.}(2022)\citenamefont
  {Trevisan}, \citenamefont {Arribi}, \citenamefont {Heinonen}, \citenamefont
  {Slager},\ and\ \citenamefont {Orth}}]{Bicircular}%
  \BibitemOpen
  \bibfield  {author} {\bibinfo {author} {\bibfnamefont {T.~V.}\ \bibnamefont
  {Trevisan}}, \bibinfo {author} {\bibfnamefont {P.~V.}\ \bibnamefont
  {Arribi}}, \bibinfo {author} {\bibfnamefont {O.}~\bibnamefont {Heinonen}},
  \bibinfo {author} {\bibfnamefont {R.-J.}\ \bibnamefont {Slager}},\ and\
  \bibinfo {author} {\bibfnamefont {P.~P.}\ \bibnamefont {Orth}},\ }\href
  {https://doi.org/10.1103/PhysRevLett.128.066602} {\bibfield  {journal}
  {\bibinfo  {journal} {Phys. Rev. Lett.}\ }\textbf {\bibinfo {volume} {128}},\
  \bibinfo {pages} {066602} (\bibinfo {year} {2022})}\BibitemShut {NoStop}%
\bibitem [{\citenamefont {Ke}\ \emph {et~al.}(2020)\citenamefont {Ke},
  \citenamefont {Asmar},\ and\ \citenamefont {Tse}}]{Asmar2020}%
  \BibitemOpen
  \bibfield  {author} {\bibinfo {author} {\bibfnamefont {M.}~\bibnamefont
  {Ke}}, \bibinfo {author} {\bibfnamefont {M.~M.}\ \bibnamefont {Asmar}},\ and\
  \bibinfo {author} {\bibfnamefont {W.-K.}\ \bibnamefont {Tse}},\ }\href
  {https://doi.org/10.1103/PhysRevResearch.2.033228} {\bibfield  {journal}
  {\bibinfo  {journal} {Phys. Rev. Research}\ }\textbf {\bibinfo {volume}
  {2}},\ \bibinfo {pages} {033228} (\bibinfo {year} {2020})}\BibitemShut
  {NoStop}%
\bibitem [{\citenamefont {Asmar}\ and\ \citenamefont {Tse}(2021)}]{Asmar2021}%
  \BibitemOpen
  \bibfield  {author} {\bibinfo {author} {\bibfnamefont {M.~M.}\ \bibnamefont
  {Asmar}}\ and\ \bibinfo {author} {\bibfnamefont {W.-K.}\ \bibnamefont
  {Tse}},\ }\href {https://doi.org/10.1088/1367-2630/ac3efe} {\bibfield
  {journal} {\bibinfo  {journal} {New Journal of Physics}\ }\textbf {\bibinfo
  {volume} {23}},\ \bibinfo {pages} {123031} (\bibinfo {year}
  {2021})}\BibitemShut {NoStop}%
\bibitem [{\citenamefont {Najmaie}\ \emph {et~al.}(2003)\citenamefont
  {Najmaie}, \citenamefont {Bhat},\ and\ \citenamefont {Sipe}}]{spininj}%
  \BibitemOpen
  \bibfield  {author} {\bibinfo {author} {\bibfnamefont {A.}~\bibnamefont
  {Najmaie}}, \bibinfo {author} {\bibfnamefont {R.~D.~R.}\ \bibnamefont
  {Bhat}},\ and\ \bibinfo {author} {\bibfnamefont {J.~E.}\ \bibnamefont
  {Sipe}},\ }\href {https://doi.org/10.1103/PhysRevB.68.165348} {\bibfield
  {journal} {\bibinfo  {journal} {Phys. Rev. B}\ }\textbf {\bibinfo {volume}
  {68}},\ \bibinfo {pages} {165348} (\bibinfo {year} {2003})}\BibitemShut
  {NoStop}%
\bibitem [{\citenamefont {Lee}\ and\ \citenamefont {Tse}(2019)}]{FloqTunn4}%
  \BibitemOpen
  \bibfield  {author} {\bibinfo {author} {\bibfnamefont {W.-R.}\ \bibnamefont
  {Lee}}\ and\ \bibinfo {author} {\bibfnamefont {W.-K.}\ \bibnamefont {Tse}},\
  }\href {https://doi.org/10.1103/PhysRevB.99.201403} {\bibfield  {journal}
  {\bibinfo  {journal} {Phys. Rev. B}\ }\textbf {\bibinfo {volume} {99}},\
  \bibinfo {pages} {201403} (\bibinfo {year} {2019})}\BibitemShut {NoStop}%
\bibitem [{\citenamefont {Kouwenhoven}\ \emph {et~al.}(1994)\citenamefont
  {Kouwenhoven}, \citenamefont {Jauhar}, \citenamefont {McCormick},
  \citenamefont {Dixon}, \citenamefont {McEuen}, \citenamefont {Nazarov},
  \citenamefont {van~der Vaart},\ and\ \citenamefont {Foxon}}]{FloqTunn1}%
  \BibitemOpen
  \bibfield  {author} {\bibinfo {author} {\bibfnamefont {L.~P.}\ \bibnamefont
  {Kouwenhoven}}, \bibinfo {author} {\bibfnamefont {S.}~\bibnamefont {Jauhar}},
  \bibinfo {author} {\bibfnamefont {K.}~\bibnamefont {McCormick}}, \bibinfo
  {author} {\bibfnamefont {D.}~\bibnamefont {Dixon}}, \bibinfo {author}
  {\bibfnamefont {P.~L.}\ \bibnamefont {McEuen}}, \bibinfo {author}
  {\bibfnamefont {Y.~V.}\ \bibnamefont {Nazarov}}, \bibinfo {author}
  {\bibfnamefont {N.~C.}\ \bibnamefont {van~der Vaart}},\ and\ \bibinfo
  {author} {\bibfnamefont {C.~T.}\ \bibnamefont {Foxon}},\ }\href
  {https://doi.org/10.1103/PhysRevB.50.2019} {\bibfield  {journal} {\bibinfo
  {journal} {Phys. Rev. B}\ }\textbf {\bibinfo {volume} {50}},\ \bibinfo
  {pages} {2019} (\bibinfo {year} {1994})}\BibitemShut {NoStop}%
\bibitem [{\citenamefont {McIver}\ \emph {et~al.}(2020)\citenamefont {McIver},
  \citenamefont {Schulte}, \citenamefont {Stein}, \citenamefont {Matsuyama},
  \citenamefont {Jotzu}, \citenamefont {Meier},\ and\ \citenamefont
  {Cavalleri}}]{FloqExp3}%
  \BibitemOpen
  \bibfield  {author} {\bibinfo {author} {\bibfnamefont {J.~W.}\ \bibnamefont
  {McIver}}, \bibinfo {author} {\bibfnamefont {B.}~\bibnamefont {Schulte}},
  \bibinfo {author} {\bibfnamefont {F.-U.}\ \bibnamefont {Stein}}, \bibinfo
  {author} {\bibfnamefont {T.}~\bibnamefont {Matsuyama}}, \bibinfo {author}
  {\bibfnamefont {G.}~\bibnamefont {Jotzu}}, \bibinfo {author} {\bibfnamefont
  {G.}~\bibnamefont {Meier}},\ and\ \bibinfo {author} {\bibfnamefont
  {A.}~\bibnamefont {Cavalleri}},\ }\href
  {https://doi.org/10.1038/s41567-019-0698-y} {\bibfield  {journal} {\bibinfo
  {journal} {Nature physics}\ }\textbf {\bibinfo {volume} {16}},\ \bibinfo
  {pages} {38} (\bibinfo {year} {2020})}\BibitemShut {NoStop}%
\bibitem [{\citenamefont {Mahmood}\ \emph {et~al.}(2016)\citenamefont
  {Mahmood}, \citenamefont {Chan}, \citenamefont {Alpichshev}, \citenamefont
  {Gardner}, \citenamefont {Lee}, \citenamefont {Lee},\ and\ \citenamefont
  {Gedik}}]{FloqExp1}%
  \BibitemOpen
  \bibfield  {author} {\bibinfo {author} {\bibfnamefont {F.}~\bibnamefont
  {Mahmood}}, \bibinfo {author} {\bibfnamefont {C.-K.}\ \bibnamefont {Chan}},
  \bibinfo {author} {\bibfnamefont {Z.}~\bibnamefont {Alpichshev}}, \bibinfo
  {author} {\bibfnamefont {D.}~\bibnamefont {Gardner}}, \bibinfo {author}
  {\bibfnamefont {Y.}~\bibnamefont {Lee}}, \bibinfo {author} {\bibfnamefont
  {P.~A.}\ \bibnamefont {Lee}},\ and\ \bibinfo {author} {\bibfnamefont
  {N.}~\bibnamefont {Gedik}},\ }\href {https://doi.org/10.1038/nphys3609}
  {\bibfield  {journal} {\bibinfo  {journal} {Nature Physics}\ }\textbf
  {\bibinfo {volume} {12}},\ \bibinfo {pages} {306} (\bibinfo {year}
  {2016})}\BibitemShut {NoStop}%
\bibitem [{\citenamefont {Wang}\ \emph {et~al.}(2013)\citenamefont {Wang},
  \citenamefont {Steinberg}, \citenamefont {Jarillo-Herrero},\ and\
  \citenamefont {Gedik}}]{FloqExp2}%
  \BibitemOpen
  \bibfield  {author} {\bibinfo {author} {\bibfnamefont {Y.~H.}\ \bibnamefont
  {Wang}}, \bibinfo {author} {\bibfnamefont {H.}~\bibnamefont {Steinberg}},
  \bibinfo {author} {\bibfnamefont {P.}~\bibnamefont {Jarillo-Herrero}},\ and\
  \bibinfo {author} {\bibfnamefont {N.}~\bibnamefont {Gedik}},\ }\href
  {https://doi.org/10.1126/science.1239834} {\bibfield  {journal} {\bibinfo
  {journal} {Science}\ }\textbf {\bibinfo {volume} {342}},\ \bibinfo {pages}
  {453} (\bibinfo {year} {2013})}\BibitemShut {NoStop}%
\bibitem [{\citenamefont {Kittel}(1969)}]{kittelbook}%
  \BibitemOpen
  \bibfield  {author} {\bibinfo {author} {\bibfnamefont {C.}~\bibnamefont
  {Kittel}},\ }\href {https://doi.org/10.1016/S0081-1947(08)60030-2} {\emph
  {\bibinfo {title} {Indirect Exchange Interactions in Metals}}},\ edited by\
  \bibinfo {editor} {\bibfnamefont {F.}~\bibnamefont {Seitz}}, \bibinfo
  {editor} {\bibfnamefont {D.}~\bibnamefont {Turnbull}},\ and\ \bibinfo
  {editor} {\bibfnamefont {H.}~\bibnamefont {Ehrenreich}},\ \bibinfo {series}
  {Solid State Physics}, Vol.~\bibinfo {volume} {22}\ (\bibinfo  {publisher}
  {Academic Press},\ \bibinfo {address} {New York,\;},\ \bibinfo {year}
  {1969})\ pp.\ \bibinfo {pages} {1 -- 26}\BibitemShut {NoStop}%
\bibitem [{\citenamefont {Black-Schaffer}(2010)}]{Graphenerkky}%
  \BibitemOpen
  \bibfield  {author} {\bibinfo {author} {\bibfnamefont {A.~M.}\ \bibnamefont
  {Black-Schaffer}},\ }\href {https://doi.org/10.1103/PhysRevB.81.205416}
  {\bibfield  {journal} {\bibinfo  {journal} {Phys. Rev. B}\ }\textbf {\bibinfo
  {volume} {81}},\ \bibinfo {pages} {205416} (\bibinfo {year}
  {2010})}\BibitemShut {NoStop}%
\bibitem [{\citenamefont {Sherafati}\ and\ \citenamefont
  {Satpathy}(2011)}]{RKKYintgraph}%
  \BibitemOpen
  \bibfield  {author} {\bibinfo {author} {\bibfnamefont {M.}~\bibnamefont
  {Sherafati}}\ and\ \bibinfo {author} {\bibfnamefont {S.}~\bibnamefont
  {Satpathy}},\ }\href {https://doi.org/10.1103/PhysRevB.83.165425} {\bibfield
  {journal} {\bibinfo  {journal} {Phys. Rev. B}\ }\textbf {\bibinfo {volume}
  {83}},\ \bibinfo {pages} {165425} (\bibinfo {year} {2011})}\BibitemShut
  {NoStop}%
\bibitem [{\citenamefont {Hatami}\ \emph {et~al.}(2014)\citenamefont {Hatami},
  \citenamefont {Kernreiter},\ and\ \citenamefont {Z\"ulicke}}]{MOS2RKKY}%
  \BibitemOpen
  \bibfield  {author} {\bibinfo {author} {\bibfnamefont {H.}~\bibnamefont
  {Hatami}}, \bibinfo {author} {\bibfnamefont {T.}~\bibnamefont {Kernreiter}},\
  and\ \bibinfo {author} {\bibfnamefont {U.}~\bibnamefont {Z\"ulicke}},\ }\href
  {https://doi.org/10.1103/PhysRevB.90.045412} {\bibfield  {journal} {\bibinfo
  {journal} {Phys. Rev. B}\ }\textbf {\bibinfo {volume} {90}},\ \bibinfo
  {pages} {045412} (\bibinfo {year} {2014})}\BibitemShut {NoStop}%
\bibitem [{\citenamefont {Newhouse-Illige}\ \emph {et~al.}(2017)\citenamefont
  {Newhouse-Illige}, \citenamefont {Liu}, \citenamefont {Xu}, \citenamefont
  {Hickey}, \citenamefont {Kundu}, \citenamefont {Almasi}, \citenamefont {Bi},
  \citenamefont {Wang}, \citenamefont {Freeland}, \citenamefont {Keavney},
  \citenamefont {Sun}, \citenamefont {Xu}, \citenamefont {Rosales},
  \citenamefont {Cheng}, \citenamefont {Zhang}, \citenamefont {Mkhoyan},\ and\
  \citenamefont {Wang}}]{RKKYvolt3}%
  \BibitemOpen
  \bibfield  {author} {\bibinfo {author} {\bibfnamefont {T.}~\bibnamefont
  {Newhouse-Illige}}, \bibinfo {author} {\bibfnamefont {Y.}~\bibnamefont
  {Liu}}, \bibinfo {author} {\bibfnamefont {M.}~\bibnamefont {Xu}}, \bibinfo
  {author} {\bibfnamefont {D.~R.}\ \bibnamefont {Hickey}}, \bibinfo {author}
  {\bibfnamefont {A.}~\bibnamefont {Kundu}}, \bibinfo {author} {\bibfnamefont
  {H.}~\bibnamefont {Almasi}}, \bibinfo {author} {\bibfnamefont
  {C.}~\bibnamefont {Bi}}, \bibinfo {author} {\bibfnamefont {X.}~\bibnamefont
  {Wang}}, \bibinfo {author} {\bibfnamefont {J.~W.}\ \bibnamefont {Freeland}},
  \bibinfo {author} {\bibfnamefont {D.~J.}\ \bibnamefont {Keavney}}, \bibinfo
  {author} {\bibfnamefont {C.~J.}\ \bibnamefont {Sun}}, \bibinfo {author}
  {\bibfnamefont {Y.~H.}\ \bibnamefont {Xu}}, \bibinfo {author} {\bibfnamefont
  {M.}~\bibnamefont {Rosales}}, \bibinfo {author} {\bibfnamefont {X.~M.}\
  \bibnamefont {Cheng}}, \bibinfo {author} {\bibfnamefont {S.}~\bibnamefont
  {Zhang}}, \bibinfo {author} {\bibfnamefont {K.~A.}\ \bibnamefont {Mkhoyan}},\
  and\ \bibinfo {author} {\bibfnamefont {W.~G.}\ \bibnamefont {Wang}},\ }\href
  {https://doi.org/10.1038/ncomms15232} {\bibfield  {journal} {\bibinfo
  {journal} {Nat. Commun.}\ }\textbf {\bibinfo {volume} {8}},\ \bibinfo {pages}
  {15232} (\bibinfo {year} {2017})}\BibitemShut {NoStop}%
\bibitem [{\citenamefont {Yang}\ \emph {et~al.}(2018)\citenamefont {Yang},
  \citenamefont {Wang}, \citenamefont {Zhou}, \citenamefont {Wang},
  \citenamefont {Zhang}, \citenamefont {Zhao}, \citenamefont {Dong},
  \citenamefont {Cheng}, \citenamefont {Min}, \citenamefont {Hu}, \citenamefont
  {Chen}, \citenamefont {Xia},\ and\ \citenamefont {Liu}}]{RKKYvolt4}%
  \BibitemOpen
  \bibfield  {author} {\bibinfo {author} {\bibfnamefont {Q.}~\bibnamefont
  {Yang}}, \bibinfo {author} {\bibfnamefont {L.}~\bibnamefont {Wang}}, \bibinfo
  {author} {\bibfnamefont {Z.}~\bibnamefont {Zhou}}, \bibinfo {author}
  {\bibfnamefont {L.}~\bibnamefont {Wang}}, \bibinfo {author} {\bibfnamefont
  {Y.}~\bibnamefont {Zhang}}, \bibinfo {author} {\bibfnamefont
  {S.}~\bibnamefont {Zhao}}, \bibinfo {author} {\bibfnamefont {G.}~\bibnamefont
  {Dong}}, \bibinfo {author} {\bibfnamefont {Y.}~\bibnamefont {Cheng}},
  \bibinfo {author} {\bibfnamefont {T.}~\bibnamefont {Min}}, \bibinfo {author}
  {\bibfnamefont {Z.}~\bibnamefont {Hu}}, \bibinfo {author} {\bibfnamefont
  {W.}~\bibnamefont {Chen}}, \bibinfo {author} {\bibfnamefont {K.}~\bibnamefont
  {Xia}},\ and\ \bibinfo {author} {\bibfnamefont {M.}~\bibnamefont {Liu}},\
  }\href {https://doi.org/10.1038/s41467-018-03356-z} {\bibfield  {journal}
  {\bibinfo  {journal} {Nat. Commun.}\ }\textbf {\bibinfo {volume} {9}},\
  \bibinfo {pages} {991} (\bibinfo {year} {2018})}\BibitemShut {NoStop}%
\bibitem [{\citenamefont {Unguris}\ \emph {et~al.}(1991)\citenamefont
  {Unguris}, \citenamefont {Celotta},\ and\ \citenamefont {Pierce}}]{exp1}%
  \BibitemOpen
  \bibfield  {author} {\bibinfo {author} {\bibfnamefont {J.}~\bibnamefont
  {Unguris}}, \bibinfo {author} {\bibfnamefont {R.~J.}\ \bibnamefont
  {Celotta}},\ and\ \bibinfo {author} {\bibfnamefont {D.~T.}\ \bibnamefont
  {Pierce}},\ }\href {https://doi.org/10.1103/PhysRevLett.67.140} {\bibfield
  {journal} {\bibinfo  {journal} {Phys. Rev. Lett.}\ }\textbf {\bibinfo
  {volume} {67}},\ \bibinfo {pages} {140} (\bibinfo {year} {1991})}\BibitemShut
  {NoStop}%
\bibitem [{\citenamefont {Purcell}\ \emph {et~al.}(1991)\citenamefont
  {Purcell}, \citenamefont {Folkerts}, \citenamefont {Johnson}, \citenamefont
  {McGee}, \citenamefont {Jager}, \citenamefont {aan~de Stegge}, \citenamefont
  {Zeper}, \citenamefont {Hoving},\ and\ \citenamefont {Gr\"unberg}}]{exp2}%
  \BibitemOpen
  \bibfield  {author} {\bibinfo {author} {\bibfnamefont {S.~T.}\ \bibnamefont
  {Purcell}}, \bibinfo {author} {\bibfnamefont {W.}~\bibnamefont {Folkerts}},
  \bibinfo {author} {\bibfnamefont {M.~T.}\ \bibnamefont {Johnson}}, \bibinfo
  {author} {\bibfnamefont {N.~W.~E.}\ \bibnamefont {McGee}}, \bibinfo {author}
  {\bibfnamefont {K.}~\bibnamefont {Jager}}, \bibinfo {author} {\bibfnamefont
  {J.}~\bibnamefont {aan~de Stegge}}, \bibinfo {author} {\bibfnamefont {W.~B.}\
  \bibnamefont {Zeper}}, \bibinfo {author} {\bibfnamefont {W.}~\bibnamefont
  {Hoving}},\ and\ \bibinfo {author} {\bibfnamefont {P.}~\bibnamefont
  {Gr\"unberg}},\ }\href {https://doi.org/10.1103/PhysRevLett.67.903}
  {\bibfield  {journal} {\bibinfo  {journal} {Phys. Rev. Lett.}\ }\textbf
  {\bibinfo {volume} {67}},\ \bibinfo {pages} {903} (\bibinfo {year}
  {1991})}\BibitemShut {NoStop}%
\bibitem [{\citenamefont {Bruno}\ and\ \citenamefont
  {Chappert}(1991)}]{LayersRKKY1}%
  \BibitemOpen
  \bibfield  {author} {\bibinfo {author} {\bibfnamefont {P.}~\bibnamefont
  {Bruno}}\ and\ \bibinfo {author} {\bibfnamefont {C.}~\bibnamefont
  {Chappert}},\ }\href {https://doi.org/10.1103/PhysRevLett.67.1602} {\bibfield
   {journal} {\bibinfo  {journal} {Phys. Rev. Lett.}\ }\textbf {\bibinfo
  {volume} {67}},\ \bibinfo {pages} {1602} (\bibinfo {year}
  {1991})}\BibitemShut {NoStop}%
\bibitem [{\citenamefont {Harper}\ \emph {et~al.}(2020)\citenamefont {Harper},
  \citenamefont {Roy}, \citenamefont {Rudner},\ and\ \citenamefont
  {Sondhi}}]{flreview4}%
  \BibitemOpen
  \bibfield  {author} {\bibinfo {author} {\bibfnamefont {F.}~\bibnamefont
  {Harper}}, \bibinfo {author} {\bibfnamefont {R.}~\bibnamefont {Roy}},
  \bibinfo {author} {\bibfnamefont {M.~S.}\ \bibnamefont {Rudner}},\ and\
  \bibinfo {author} {\bibfnamefont {S.}~\bibnamefont {Sondhi}},\ }\href
  {https://doi.org/10.1146/annurev-conmatphys-031218-013721} {\bibfield
  {journal} {\bibinfo  {journal} {Annual Review of Condensed Matter Physics}\
  }\textbf {\bibinfo {volume} {11}},\ \bibinfo {pages} {345} (\bibinfo {year}
  {2020})}\BibitemShut {NoStop}%
\bibitem [{\citenamefont {Ono}\ and\ \citenamefont {Ishihara}(2019)}]{ono}%
  \BibitemOpen
  \bibfield  {author} {\bibinfo {author} {\bibfnamefont {A.}~\bibnamefont
  {Ono}}\ and\ \bibinfo {author} {\bibfnamefont {S.}~\bibnamefont {Ishihara}},\
  }\href {https://doi.org/10.1103/PhysRevB.100.075127} {\bibfield  {journal}
  {\bibinfo  {journal} {Phys. Rev. B}\ }\textbf {\bibinfo {volume} {100}},\
  \bibinfo {pages} {075127} (\bibinfo {year} {2019})}\BibitemShut {NoStop}%
\bibitem [{\citenamefont {Bloembergen}\ and\ \citenamefont
  {Rowland}(1955)}]{BR}%
  \BibitemOpen
  \bibfield  {author} {\bibinfo {author} {\bibfnamefont {N.}~\bibnamefont
  {Bloembergen}}\ and\ \bibinfo {author} {\bibfnamefont {T.~J.}\ \bibnamefont
  {Rowland}},\ }\href {https://doi.org/10.1103/PhysRev.97.1679} {\bibfield
  {journal} {\bibinfo  {journal} {Phys. Rev.}\ }\textbf {\bibinfo {volume}
  {97}},\ \bibinfo {pages} {1679} (\bibinfo {year} {1955})}\BibitemShut
  {NoStop}%
\bibitem [{\citenamefont {Joshi}(2016)}]{Spintronics1}%
  \BibitemOpen
  \bibfield  {author} {\bibinfo {author} {\bibfnamefont {V.~K.}\ \bibnamefont
  {Joshi}},\ }\href
  {https://doi.org/https://doi.org/10.1016/j.jestch.2016.05.002} {\bibfield
  {journal} {\bibinfo  {journal} {Engineering Science and Technology, an
  International Journal}\ }\textbf {\bibinfo {volume} {19}},\ \bibinfo {pages}
  {1503 } (\bibinfo {year} {2016})}\BibitemShut {NoStop}%
\bibitem [{\citenamefont {Finocchio}\ \emph {et~al.}(2016)\citenamefont
  {Finocchio}, \citenamefont {Büttner}, \citenamefont {Tomasello},
  \citenamefont {Carpentieri},\ and\ \citenamefont {Kläui}}]{skyrmions}%
  \BibitemOpen
  \bibfield  {author} {\bibinfo {author} {\bibfnamefont {G.}~\bibnamefont
  {Finocchio}}, \bibinfo {author} {\bibfnamefont {F.}~\bibnamefont {Büttner}},
  \bibinfo {author} {\bibfnamefont {R.}~\bibnamefont {Tomasello}}, \bibinfo
  {author} {\bibfnamefont {M.}~\bibnamefont {Carpentieri}},\ and\ \bibinfo
  {author} {\bibfnamefont {M.}~\bibnamefont {Kläui}},\ }\href
  {http://stacks.iop.org/0022-3727/49/i=42/a=423001} {\bibfield  {journal}
  {\bibinfo  {journal} {Journal of Physics D: Applied Physics}\ }\textbf
  {\bibinfo {volume} {49}},\ \bibinfo {pages} {423001} (\bibinfo {year}
  {2016})}\BibitemShut {NoStop}%
\bibitem [{\citenamefont {Garst}\ \emph {et~al.}(2017)\citenamefont {Garst},
  \citenamefont {Waizner},\ and\ \citenamefont {Grundler}}]{helix}%
  \BibitemOpen
  \bibfield  {author} {\bibinfo {author} {\bibfnamefont {M.}~\bibnamefont
  {Garst}}, \bibinfo {author} {\bibfnamefont {J.}~\bibnamefont {Waizner}},\
  and\ \bibinfo {author} {\bibfnamefont {D.}~\bibnamefont {Grundler}},\ }\href
  {https://doi.org/10.1088/1361-6463/aa7573} {\bibfield  {journal} {\bibinfo
  {journal} {Journal of Physics D: Applied Physics}\ }\textbf {\bibinfo
  {volume} {50}},\ \bibinfo {pages} {293002} (\bibinfo {year}
  {2017})}\BibitemShut {NoStop}%
\bibitem [{\citenamefont {Je}\ \emph {et~al.}(2013)\citenamefont {Je},
  \citenamefont {Kim}, \citenamefont {Yoo}, \citenamefont {Min}, \citenamefont
  {Lee},\ and\ \citenamefont {Choe}}]{domainwall}%
  \BibitemOpen
  \bibfield  {author} {\bibinfo {author} {\bibfnamefont {S.-G.}\ \bibnamefont
  {Je}}, \bibinfo {author} {\bibfnamefont {D.-H.}\ \bibnamefont {Kim}},
  \bibinfo {author} {\bibfnamefont {S.-C.}\ \bibnamefont {Yoo}}, \bibinfo
  {author} {\bibfnamefont {B.-C.}\ \bibnamefont {Min}}, \bibinfo {author}
  {\bibfnamefont {K.-J.}\ \bibnamefont {Lee}},\ and\ \bibinfo {author}
  {\bibfnamefont {S.-B.}\ \bibnamefont {Choe}},\ }\href
  {https://doi.org/10.1103/PhysRevB.88.214401} {\bibfield  {journal} {\bibinfo
  {journal} {Phys. Rev. B}\ }\textbf {\bibinfo {volume} {88}},\ \bibinfo
  {pages} {214401} (\bibinfo {year} {2013})}\BibitemShut {NoStop}%
\bibitem [{\citenamefont {Dzyaloshinsky}(1958)}]{DM1}%
  \BibitemOpen
  \bibfield  {author} {\bibinfo {author} {\bibfnamefont {I.}~\bibnamefont
  {Dzyaloshinsky}},\ }\href
  {https://doi.org/https://doi.org/10.1016/0022-3697(58)90076-3} {\bibfield
  {journal} {\bibinfo  {journal} {Journal of Physics and Chemistry of Solids}\
  }\textbf {\bibinfo {volume} {4}},\ \bibinfo {pages} {241 } (\bibinfo {year}
  {1958})}\BibitemShut {NoStop}%
\bibitem [{\citenamefont {Moriya}(1960)}]{DM2}%
  \BibitemOpen
  \bibfield  {author} {\bibinfo {author} {\bibfnamefont {T.}~\bibnamefont
  {Moriya}},\ }\href {https://doi.org/10.1103/PhysRev.120.91} {\bibfield
  {journal} {\bibinfo  {journal} {Phys. Rev.}\ }\textbf {\bibinfo {volume}
  {120}},\ \bibinfo {pages} {91} (\bibinfo {year} {1960})}\BibitemShut
  {NoStop}%
\bibitem [{\citenamefont {Yang}\ \emph {et~al.}(2013)\citenamefont {Yang},
  \citenamefont {Dolev}, \citenamefont {Zhang}, \citenamefont {Zhao},
  \citenamefont {Fried}, \citenamefont {Schemm}, \citenamefont {Liu},
  \citenamefont {Palevski}, \citenamefont {Marshall}, \citenamefont {Risbud},\
  and\ \citenamefont {Kapitulnik}}]{EuS2}%
  \BibitemOpen
  \bibfield  {author} {\bibinfo {author} {\bibfnamefont {Q.~I.}\ \bibnamefont
  {Yang}}, \bibinfo {author} {\bibfnamefont {M.}~\bibnamefont {Dolev}},
  \bibinfo {author} {\bibfnamefont {L.}~\bibnamefont {Zhang}}, \bibinfo
  {author} {\bibfnamefont {J.}~\bibnamefont {Zhao}}, \bibinfo {author}
  {\bibfnamefont {A.~D.}\ \bibnamefont {Fried}}, \bibinfo {author}
  {\bibfnamefont {E.}~\bibnamefont {Schemm}}, \bibinfo {author} {\bibfnamefont
  {M.}~\bibnamefont {Liu}}, \bibinfo {author} {\bibfnamefont {A.}~\bibnamefont
  {Palevski}}, \bibinfo {author} {\bibfnamefont {A.~F.}\ \bibnamefont
  {Marshall}}, \bibinfo {author} {\bibfnamefont {S.~H.}\ \bibnamefont
  {Risbud}},\ and\ \bibinfo {author} {\bibfnamefont {A.}~\bibnamefont
  {Kapitulnik}},\ }\href {https://doi.org/10.1103/PhysRevB.88.081407}
  {\bibfield  {journal} {\bibinfo  {journal} {Phys. Rev. B}\ }\textbf {\bibinfo
  {volume} {88}},\ \bibinfo {pages} {081407} (\bibinfo {year}
  {2013})}\BibitemShut {NoStop}%
\bibitem [{\citenamefont {Eremeev}\ \emph {et~al.}(2015)\citenamefont
  {Eremeev}, \citenamefont {Men}, \citenamefont {Tugushev}, \citenamefont
  {Chulkov} \emph {et~al.}}]{EuS3}%
  \BibitemOpen
  \bibfield  {author} {\bibinfo {author} {\bibfnamefont {S.}~\bibnamefont
  {Eremeev}}, \bibinfo {author} {\bibfnamefont {V.}~\bibnamefont {Men}},
  \bibinfo {author} {\bibfnamefont {V.}~\bibnamefont {Tugushev}}, \bibinfo
  {author} {\bibfnamefont {E.~V.}\ \bibnamefont {Chulkov}}, \emph {et~al.},\
  }\href {https://doi.org/https://doi.org/10.1016/j.jmmm.2014.09.029}
  {\bibfield  {journal} {\bibinfo  {journal} {Journal of Magnetism and Magnetic
  Materials}\ }\textbf {\bibinfo {volume} {383}},\ \bibinfo {pages} {30}
  (\bibinfo {year} {2015})}\BibitemShut {NoStop}%
\bibitem [{\citenamefont {Lee}\ \emph {et~al.}(2016)\citenamefont {Lee},
  \citenamefont {Katmis}, \citenamefont {Jarillo-Herrero}, \citenamefont
  {Moodera},\ and\ \citenamefont {Gedik}}]{EuS4}%
  \BibitemOpen
  \bibfield  {author} {\bibinfo {author} {\bibfnamefont {C.}~\bibnamefont
  {Lee}}, \bibinfo {author} {\bibfnamefont {F.}~\bibnamefont {Katmis}},
  \bibinfo {author} {\bibfnamefont {P.}~\bibnamefont {Jarillo-Herrero}},
  \bibinfo {author} {\bibfnamefont {J.~S.}\ \bibnamefont {Moodera}},\ and\
  \bibinfo {author} {\bibfnamefont {N.}~\bibnamefont {Gedik}},\ }\href
  {https://doi.org/https://doi.org/10.1038/ncomms12014} {\bibfield  {journal}
  {\bibinfo  {journal} {Nat. Commun.}\ }\textbf {\bibinfo {volume} {7}},\
  \bibinfo {pages} {1} (\bibinfo {year} {2016})}\BibitemShut {NoStop}%
\bibitem [{\citenamefont {Bahramy}\ and\ \citenamefont {Ogawa}(2017)}]{BiTeI1}%
  \BibitemOpen
  \bibfield  {author} {\bibinfo {author} {\bibfnamefont {M.~S.}\ \bibnamefont
  {Bahramy}}\ and\ \bibinfo {author} {\bibfnamefont {N.}~\bibnamefont
  {Ogawa}},\ }\href {https://doi.org/10.1002/adma.201605911} {\bibfield
  {journal} {\bibinfo  {journal} {Advanced Materials}\ }\textbf {\bibinfo
  {volume} {29}},\ \bibinfo {pages} {1605911} (\bibinfo {year}
  {2017})}\BibitemShut {NoStop}%
\bibitem [{\citenamefont {Ishizaka}\ \emph {et~al.}(2011)\citenamefont
  {Ishizaka}, \citenamefont {Bahramy}, \citenamefont {Murakawa}, \citenamefont
  {Sakano}, \citenamefont {Shimojima}, \citenamefont {Sonobe}, \citenamefont
  {Koizumi}, \citenamefont {Shin}, \citenamefont {Miyahara}, \citenamefont
  {Kimura}, \citenamefont {Miyamoto}, \citenamefont {Okuda}, \citenamefont
  {Namatame}, \citenamefont {Taniguchi}, \citenamefont {Arita}, \citenamefont
  {Nagaosa}, \citenamefont {Kobayashi}, \citenamefont {Murakami}, \citenamefont
  {Kumai}, \citenamefont {Kaneko},\ and\ \citenamefont {Onose}}]{BiTeI2}%
  \BibitemOpen
  \bibfield  {author} {\bibinfo {author} {\bibfnamefont {K.}~\bibnamefont
  {Ishizaka}}, \bibinfo {author} {\bibfnamefont {M.~S.}\ \bibnamefont
  {Bahramy}}, \bibinfo {author} {\bibfnamefont {H.}~\bibnamefont {Murakawa}},
  \bibinfo {author} {\bibfnamefont {M.}~\bibnamefont {Sakano}}, \bibinfo
  {author} {\bibfnamefont {T.}~\bibnamefont {Shimojima}}, \bibinfo {author}
  {\bibfnamefont {T.}~\bibnamefont {Sonobe}}, \bibinfo {author} {\bibfnamefont
  {K.}~\bibnamefont {Koizumi}}, \bibinfo {author} {\bibfnamefont
  {S.}~\bibnamefont {Shin}}, \bibinfo {author} {\bibfnamefont {H.}~\bibnamefont
  {Miyahara}}, \bibinfo {author} {\bibfnamefont {A.}~\bibnamefont {Kimura}},
  \bibinfo {author} {\bibfnamefont {K.}~\bibnamefont {Miyamoto}}, \bibinfo
  {author} {\bibfnamefont {T.}~\bibnamefont {Okuda}}, \bibinfo {author}
  {\bibfnamefont {H.}~\bibnamefont {Namatame}}, \bibinfo {author}
  {\bibfnamefont {M.}~\bibnamefont {Taniguchi}}, \bibinfo {author}
  {\bibfnamefont {R.}~\bibnamefont {Arita}}, \bibinfo {author} {\bibfnamefont
  {N.}~\bibnamefont {Nagaosa}}, \bibinfo {author} {\bibfnamefont
  {K.}~\bibnamefont {Kobayashi}}, \bibinfo {author} {\bibfnamefont
  {Y.}~\bibnamefont {Murakami}}, \bibinfo {author} {\bibfnamefont
  {R.}~\bibnamefont {Kumai}}, \bibinfo {author} {\bibfnamefont
  {Y.}~\bibnamefont {Kaneko}},\ and\ \bibinfo {author} {\bibfnamefont
  {Y.}~\bibnamefont {Onose}},\ }\href {https://doi.org/10.1038/nmat3051}
  {\bibfield  {journal} {\bibinfo  {journal} {Nature Materials}\ }\textbf
  {\bibinfo {volume} {10}},\ \bibinfo {pages} {521} (\bibinfo {year}
  {2011})}\BibitemShut {NoStop}%
\bibitem [{\citenamefont {Murakawa}\ \emph {et~al.}(2013)\citenamefont
  {Murakawa}, \citenamefont {Bahramy}, \citenamefont {Tokunaga}, \citenamefont
  {Kohama}, \citenamefont {Bell}, \citenamefont {Kaneko}, \citenamefont
  {Nagaosa}, \citenamefont {Hwang},\ and\ \citenamefont {Tokura}}]{BiTeI3}%
  \BibitemOpen
  \bibfield  {author} {\bibinfo {author} {\bibfnamefont {H.}~\bibnamefont
  {Murakawa}}, \bibinfo {author} {\bibfnamefont {M.~S.}\ \bibnamefont
  {Bahramy}}, \bibinfo {author} {\bibfnamefont {M.}~\bibnamefont {Tokunaga}},
  \bibinfo {author} {\bibfnamefont {Y.}~\bibnamefont {Kohama}}, \bibinfo
  {author} {\bibfnamefont {C.}~\bibnamefont {Bell}}, \bibinfo {author}
  {\bibfnamefont {Y.}~\bibnamefont {Kaneko}}, \bibinfo {author} {\bibfnamefont
  {N.}~\bibnamefont {Nagaosa}}, \bibinfo {author} {\bibfnamefont {H.~Y.}\
  \bibnamefont {Hwang}},\ and\ \bibinfo {author} {\bibfnamefont
  {Y.}~\bibnamefont {Tokura}},\ }\href
  {https://doi.org/10.1126/science.1242247} {\bibfield  {journal} {\bibinfo
  {journal} {Science}\ }\textbf {\bibinfo {volume} {342}},\ \bibinfo {pages}
  {1490} (\bibinfo {year} {2013})}\BibitemShut {NoStop}%
\bibitem [{\citenamefont {Bahramy}\ \emph {et~al.}(2011)\citenamefont
  {Bahramy}, \citenamefont {Arita},\ and\ \citenamefont {Nagaosa}}]{BiTeI6}%
  \BibitemOpen
  \bibfield  {author} {\bibinfo {author} {\bibfnamefont {M.~S.}\ \bibnamefont
  {Bahramy}}, \bibinfo {author} {\bibfnamefont {R.}~\bibnamefont {Arita}},\
  and\ \bibinfo {author} {\bibfnamefont {N.}~\bibnamefont {Nagaosa}},\ }\href
  {https://doi.org/10.1103/PhysRevB.84.041202} {\bibfield  {journal} {\bibinfo
  {journal} {Phys. Rev. B}\ }\textbf {\bibinfo {volume} {84}},\ \bibinfo
  {pages} {041202} (\bibinfo {year} {2011})}\BibitemShut {NoStop}%
\bibitem [{\citenamefont {Bell}\ \emph {et~al.}(2013)\citenamefont {Bell},
  \citenamefont {Bahramy}, \citenamefont {Murakawa}, \citenamefont
  {Checkelsky}, \citenamefont {Arita}, \citenamefont {Kaneko}, \citenamefont
  {Onose}, \citenamefont {Tokunaga}, \citenamefont {Kohama}, \citenamefont
  {Nagaosa}, \citenamefont {Tokura},\ and\ \citenamefont
  {Hwang}}]{TransportBiTeI}%
  \BibitemOpen
  \bibfield  {author} {\bibinfo {author} {\bibfnamefont {C.}~\bibnamefont
  {Bell}}, \bibinfo {author} {\bibfnamefont {M.~S.}\ \bibnamefont {Bahramy}},
  \bibinfo {author} {\bibfnamefont {H.}~\bibnamefont {Murakawa}}, \bibinfo
  {author} {\bibfnamefont {J.~G.}\ \bibnamefont {Checkelsky}}, \bibinfo
  {author} {\bibfnamefont {R.}~\bibnamefont {Arita}}, \bibinfo {author}
  {\bibfnamefont {Y.}~\bibnamefont {Kaneko}}, \bibinfo {author} {\bibfnamefont
  {Y.}~\bibnamefont {Onose}}, \bibinfo {author} {\bibfnamefont
  {M.}~\bibnamefont {Tokunaga}}, \bibinfo {author} {\bibfnamefont
  {Y.}~\bibnamefont {Kohama}}, \bibinfo {author} {\bibfnamefont
  {N.}~\bibnamefont {Nagaosa}}, \bibinfo {author} {\bibfnamefont
  {Y.}~\bibnamefont {Tokura}},\ and\ \bibinfo {author} {\bibfnamefont {H.~Y.}\
  \bibnamefont {Hwang}},\ }\href {https://doi.org/10.1103/PhysRevB.87.081109}
  {\bibfield  {journal} {\bibinfo  {journal} {Phys. Rev. B}\ }\textbf {\bibinfo
  {volume} {87}},\ \bibinfo {pages} {081109} (\bibinfo {year}
  {2013})}\BibitemShut {NoStop}%
\bibitem [{\citenamefont {Lee}\ \emph {et~al.}(2011)\citenamefont {Lee},
  \citenamefont {Schober}, \citenamefont {Bahramy}, \citenamefont {Murakawa},
  \citenamefont {Onose}, \citenamefont {Arita}, \citenamefont {Nagaosa},\ and\
  \citenamefont {Tokura}}]{OpticalBiTeI}%
  \BibitemOpen
  \bibfield  {author} {\bibinfo {author} {\bibfnamefont {J.~S.}\ \bibnamefont
  {Lee}}, \bibinfo {author} {\bibfnamefont {G.~A.~H.}\ \bibnamefont {Schober}},
  \bibinfo {author} {\bibfnamefont {M.~S.}\ \bibnamefont {Bahramy}}, \bibinfo
  {author} {\bibfnamefont {H.}~\bibnamefont {Murakawa}}, \bibinfo {author}
  {\bibfnamefont {Y.}~\bibnamefont {Onose}}, \bibinfo {author} {\bibfnamefont
  {R.}~\bibnamefont {Arita}}, \bibinfo {author} {\bibfnamefont
  {N.}~\bibnamefont {Nagaosa}},\ and\ \bibinfo {author} {\bibfnamefont
  {Y.}~\bibnamefont {Tokura}},\ }\href
  {https://doi.org/10.1103/PhysRevLett.107.117401} {\bibfield  {journal}
  {\bibinfo  {journal} {Phys. Rev. Lett.}\ }\textbf {\bibinfo {volume} {107}},\
  \bibinfo {pages} {117401} (\bibinfo {year} {2011})}\BibitemShut {NoStop}%
\bibitem [{\citenamefont {Stephanovich}\ \emph {et~al.}(2017)\citenamefont
  {Stephanovich}, \citenamefont {Dugaev}, \citenamefont {Litvinov},\ and\
  \citenamefont {Berakdar}}]{build1}%
  \BibitemOpen
  \bibfield  {author} {\bibinfo {author} {\bibfnamefont {V.~A.}\ \bibnamefont
  {Stephanovich}}, \bibinfo {author} {\bibfnamefont {V.~K.}\ \bibnamefont
  {Dugaev}}, \bibinfo {author} {\bibfnamefont {V.~I.}\ \bibnamefont
  {Litvinov}},\ and\ \bibinfo {author} {\bibfnamefont {J.}~\bibnamefont
  {Berakdar}},\ }\href {https://doi.org/10.1103/PhysRevB.95.045307} {\bibfield
  {journal} {\bibinfo  {journal} {Phys. Rev. B}\ }\textbf {\bibinfo {volume}
  {95}},\ \bibinfo {pages} {045307} (\bibinfo {year} {2017})}\BibitemShut
  {NoStop}%
\bibitem [{\citenamefont {Stephanovich}\ \emph {et~al.}(2019)\citenamefont
  {Stephanovich}, \citenamefont {Kirichenko}, \citenamefont {Dugaev},
  \citenamefont {Barna\ifmmode~\acute{s}\else \'{s}\fi{}},\ and\ \citenamefont
  {Berakdar}}]{build2}%
  \BibitemOpen
  \bibfield  {author} {\bibinfo {author} {\bibfnamefont {V.~A.}\ \bibnamefont
  {Stephanovich}}, \bibinfo {author} {\bibfnamefont {E.~V.}\ \bibnamefont
  {Kirichenko}}, \bibinfo {author} {\bibfnamefont {V.~K.}\ \bibnamefont
  {Dugaev}}, \bibinfo {author} {\bibfnamefont {J.}~\bibnamefont
  {Barna\ifmmode~\acute{s}\else \'{s}\fi{}}},\ and\ \bibinfo {author}
  {\bibfnamefont {J.}~\bibnamefont {Berakdar}},\ }\href
  {https://doi.org/10.1103/PhysRevB.99.235302} {\bibfield  {journal} {\bibinfo
  {journal} {Phys. Rev. B}\ }\textbf {\bibinfo {volume} {99}},\ \bibinfo
  {pages} {235302} (\bibinfo {year} {2019})}\BibitemShut {NoStop}%
\bibitem [{not()}]{note1}%
  \BibitemOpen
  \href@noop {} {}\bibinfo {note} {Although the Rashba Hamiltonian perfectly
  describes the spectrum of electrons in BiTeI, light-induced spin-flip
  transitions associated with the Rashba coupled electrons exhibit a strong
  dependence on the electron momentum magnitude, a feature that is not present
  in the Rashba model, leading to the underestimation of the effects of these
  transitions~\cite{flip}. In the case presented in the manuscript, this type
  of transitions are absent due to the high frequency of light.}\BibitemShut
  {Stop}%
\bibitem [{\citenamefont {Shirley}(1965)}]{Shirley}%
  \BibitemOpen
  \bibfield  {author} {\bibinfo {author} {\bibfnamefont {J.~H.}\ \bibnamefont
  {Shirley}},\ }\href {https://doi.org/10.1103/PhysRev.138.B979} {\bibfield
  {journal} {\bibinfo  {journal} {Phys. Rev.}\ }\textbf {\bibinfo {volume}
  {138}},\ \bibinfo {pages} {B979} (\bibinfo {year} {1965})}\BibitemShut
  {NoStop}%
\bibitem [{\citenamefont {Eckardt}\ and\ \citenamefont
  {Anisimovas}(2015)}]{VanVleck}%
  \BibitemOpen
  \bibfield  {author} {\bibinfo {author} {\bibfnamefont {A.}~\bibnamefont
  {Eckardt}}\ and\ \bibinfo {author} {\bibfnamefont {E.}~\bibnamefont
  {Anisimovas}},\ }\href {https://doi.org/10.1088/1367-2630/17/9/093039}
  {\bibfield  {journal} {\bibinfo  {journal} {New Journal of Physics}\ }\textbf
  {\bibinfo {volume} {17}},\ \bibinfo {pages} {093039} (\bibinfo {year}
  {2015})}\BibitemShut {NoStop}%
\bibitem [{\citenamefont {Rodriguez-Vega}\ \emph {et~al.}(2021)\citenamefont
  {Rodriguez-Vega}, \citenamefont {Vogl},\ and\ \citenamefont
  {Fiete}}]{martin2}%
  \BibitemOpen
  \bibfield  {author} {\bibinfo {author} {\bibfnamefont {M.}~\bibnamefont
  {Rodriguez-Vega}}, \bibinfo {author} {\bibfnamefont {M.}~\bibnamefont
  {Vogl}},\ and\ \bibinfo {author} {\bibfnamefont {G.~A.}\ \bibnamefont
  {Fiete}},\ }\href {https://doi.org/https://doi.org/10.1016/j.aop.2021.168434}
  {\bibfield  {journal} {\bibinfo  {journal} {Annals of Physics}\ ,\ \bibinfo
  {pages} {168434}} (\bibinfo {year} {2021})}\BibitemShut {NoStop}%
\bibitem [{\citenamefont {Mikami}\ \emph {et~al.}(2016)\citenamefont {Mikami},
  \citenamefont {Kitamura}, \citenamefont {Yasuda}, \citenamefont {Tsuji},
  \citenamefont {Oka},\ and\ \citenamefont {Aoki}}]{off-res}%
  \BibitemOpen
  \bibfield  {author} {\bibinfo {author} {\bibfnamefont {T.}~\bibnamefont
  {Mikami}}, \bibinfo {author} {\bibfnamefont {S.}~\bibnamefont {Kitamura}},
  \bibinfo {author} {\bibfnamefont {K.}~\bibnamefont {Yasuda}}, \bibinfo
  {author} {\bibfnamefont {N.}~\bibnamefont {Tsuji}}, \bibinfo {author}
  {\bibfnamefont {T.}~\bibnamefont {Oka}},\ and\ \bibinfo {author}
  {\bibfnamefont {H.}~\bibnamefont {Aoki}},\ }\href
  {https://doi.org/10.1103/PhysRevB.93.144307} {\bibfield  {journal} {\bibinfo
  {journal} {Phys. Rev. B}\ }\textbf {\bibinfo {volume} {93}},\ \bibinfo
  {pages} {144307} (\bibinfo {year} {2016})}\BibitemShut {NoStop}%
\bibitem [{\citenamefont {Stiles}(2006)}]{stiles}%
  \BibitemOpen
  \bibfield  {author} {\bibinfo {author} {\bibfnamefont {M.}~\bibnamefont
  {Stiles}},\ }in\ \href
  {https://doi.org/https://doi.org/10.1016/S1572-0934(05)01003-6} {\emph
  {\bibinfo {booktitle} {Nanomagnetism: Ultrathin Films, Multilayers and
  Nanostructures}}},\ \bibinfo {series} {Contemporary Concepts of Condensed
  Matter Science}, Vol.~\bibinfo {volume} {1},\ \bibinfo {editor} {edited by\
  \bibinfo {editor} {\bibfnamefont {D.}~\bibnamefont {Mills}}\ and\ \bibinfo
  {editor} {\bibfnamefont {J.}~\bibnamefont {Bland}}}\ (\bibinfo  {publisher}
  {Elsevier},\ \bibinfo {year} {2006})\ pp.\ \bibinfo {pages} {51 --
  76}\BibitemShut {NoStop}%
\bibitem [{\citenamefont {Asmar}\ and\ \citenamefont {Tse}(2019)}]{Asmar2019}%
  \BibitemOpen
  \bibfield  {author} {\bibinfo {author} {\bibfnamefont {M.~M.}\ \bibnamefont
  {Asmar}}\ and\ \bibinfo {author} {\bibfnamefont {W.-K.}\ \bibnamefont
  {Tse}},\ }\href {https://doi.org/10.1103/PhysRevB.100.014410} {\bibfield
  {journal} {\bibinfo  {journal} {Phys. Rev. B}\ }\textbf {\bibinfo {volume}
  {100}},\ \bibinfo {pages} {014410} (\bibinfo {year} {2019})}\BibitemShut
  {NoStop}%
\bibitem [{\citenamefont {Yarmohammadi}\ \emph {et~al.}(2023)\citenamefont
  {Yarmohammadi}, \citenamefont {Bukov},\ and\ \citenamefont
  {Kolodrubetz}}]{Ideal1}%
  \BibitemOpen
  \bibfield  {author} {\bibinfo {author} {\bibfnamefont {M.}~\bibnamefont
  {Yarmohammadi}}, \bibinfo {author} {\bibfnamefont {M.}~\bibnamefont
  {Bukov}},\ and\ \bibinfo {author} {\bibfnamefont {M.~H.}\ \bibnamefont
  {Kolodrubetz}},\ }\href {https://doi.org/10.1103/PhysRevB.107.054439}
  {\bibfield  {journal} {\bibinfo  {journal} {Phys. Rev. B}\ }\textbf {\bibinfo
  {volume} {107}},\ \bibinfo {pages} {054439} (\bibinfo {year}
  {2023})}\BibitemShut {NoStop}%
\bibitem [{\citenamefont {Kumar}\ \emph
  {et~al.}(2020{\natexlab{b}})\citenamefont {Kumar}, \citenamefont
  {Rodriguez-Vega}, \citenamefont {Pereg-Barnea},\ and\ \citenamefont
  {Seradjeh}}]{Ideal2}%
  \BibitemOpen
  \bibfield  {author} {\bibinfo {author} {\bibfnamefont {A.}~\bibnamefont
  {Kumar}}, \bibinfo {author} {\bibfnamefont {M.}~\bibnamefont
  {Rodriguez-Vega}}, \bibinfo {author} {\bibfnamefont {T.}~\bibnamefont
  {Pereg-Barnea}},\ and\ \bibinfo {author} {\bibfnamefont {B.}~\bibnamefont
  {Seradjeh}},\ }\href {https://doi.org/10.1103/PhysRevB.101.174314} {\bibfield
   {journal} {\bibinfo  {journal} {Phys. Rev. B}\ }\textbf {\bibinfo {volume}
  {101}},\ \bibinfo {pages} {174314} (\bibinfo {year}
  {2020}{\natexlab{b}})}\BibitemShut {NoStop}%
\bibitem [{\citenamefont {Menon}\ \emph {et~al.}(2018)\citenamefont {Menon},
  \citenamefont {Chowdhury},\ and\ \citenamefont {Basu}}]{Ideal3}%
  \BibitemOpen
  \bibfield  {author} {\bibinfo {author} {\bibfnamefont {A.}~\bibnamefont
  {Menon}}, \bibinfo {author} {\bibfnamefont {D.}~\bibnamefont {Chowdhury}},\
  and\ \bibinfo {author} {\bibfnamefont {B.}~\bibnamefont {Basu}},\ }\href
  {https://doi.org/10.1103/PhysRevB.98.205109} {\bibfield  {journal} {\bibinfo
  {journal} {Phys. Rev. B}\ }\textbf {\bibinfo {volume} {98}},\ \bibinfo
  {pages} {205109} (\bibinfo {year} {2018})}\BibitemShut {NoStop}%
\bibitem [{\citenamefont {Seshadri}\ and\ \citenamefont
  {Pereg-Barnea}(2023)}]{Ideal4}%
  \BibitemOpen
  \bibfield  {author} {\bibinfo {author} {\bibfnamefont {R.}~\bibnamefont
  {Seshadri}}\ and\ \bibinfo {author} {\bibfnamefont {T.}~\bibnamefont
  {Pereg-Barnea}},\ }\href {https://doi.org/10.1103/PhysRevB.108.235406}
  {\bibfield  {journal} {\bibinfo  {journal} {Phys. Rev. B}\ }\textbf {\bibinfo
  {volume} {108}},\ \bibinfo {pages} {235406} (\bibinfo {year}
  {2023})}\BibitemShut {NoStop}%
\bibitem [{\citenamefont {Du}\ \emph {et~al.}(2017)\citenamefont {Du},
  \citenamefont {Zhou},\ and\ \citenamefont {Fiete}}]{Ideal5}%
  \BibitemOpen
  \bibfield  {author} {\bibinfo {author} {\bibfnamefont {L.}~\bibnamefont
  {Du}}, \bibinfo {author} {\bibfnamefont {X.}~\bibnamefont {Zhou}},\ and\
  \bibinfo {author} {\bibfnamefont {G.~A.}\ \bibnamefont {Fiete}},\ }\href
  {https://doi.org/10.1103/PhysRevB.95.035136} {\bibfield  {journal} {\bibinfo
  {journal} {Phys. Rev. B}\ }\textbf {\bibinfo {volume} {95}},\ \bibinfo
  {pages} {035136} (\bibinfo {year} {2017})}\BibitemShut {NoStop}%
\bibitem [{\citenamefont {Dehghani}\ and\ \citenamefont
  {Mitra}(2015{\natexlab{b}})}]{Ideal6}%
  \BibitemOpen
  \bibfield  {author} {\bibinfo {author} {\bibfnamefont {H.}~\bibnamefont
  {Dehghani}}\ and\ \bibinfo {author} {\bibfnamefont {A.}~\bibnamefont
  {Mitra}},\ }\href {https://doi.org/10.1103/PhysRevB.92.165111} {\bibfield
  {journal} {\bibinfo  {journal} {Phys. Rev. B}\ }\textbf {\bibinfo {volume}
  {92}},\ \bibinfo {pages} {165111} (\bibinfo {year}
  {2015}{\natexlab{b}})}\BibitemShut {NoStop}%
\bibitem [{\citenamefont {Dabiri}\ \emph {et~al.}(2022)\citenamefont {Dabiri},
  \citenamefont {Cheraghchi},\ and\ \citenamefont {Sadeghi}}]{Ideal7}%
  \BibitemOpen
  \bibfield  {author} {\bibinfo {author} {\bibfnamefont {S.~S.}\ \bibnamefont
  {Dabiri}}, \bibinfo {author} {\bibfnamefont {H.}~\bibnamefont {Cheraghchi}},\
  and\ \bibinfo {author} {\bibfnamefont {A.}~\bibnamefont {Sadeghi}},\ }\href
  {https://doi.org/10.1103/PhysRevB.106.165423} {\bibfield  {journal} {\bibinfo
   {journal} {Phys. Rev. B}\ }\textbf {\bibinfo {volume} {106}},\ \bibinfo
  {pages} {165423} (\bibinfo {year} {2022})}\BibitemShut {NoStop}%
\bibitem [{\citenamefont {Kohn}(1959)}]{kohnanomaly}%
  \BibitemOpen
  \bibfield  {author} {\bibinfo {author} {\bibfnamefont {W.}~\bibnamefont
  {Kohn}},\ }\href {https://doi.org/10.1103/PhysRevLett.2.393} {\bibfield
  {journal} {\bibinfo  {journal} {Phys. Rev. Lett.}\ }\textbf {\bibinfo
  {volume} {2}},\ \bibinfo {pages} {393} (\bibinfo {year} {1959})}\BibitemShut
  {NoStop}%
\bibitem [{\citenamefont {Bruno}\ and\ \citenamefont
  {Chappert}(1992)}]{LayersRKKY2}%
  \BibitemOpen
  \bibfield  {author} {\bibinfo {author} {\bibfnamefont {P.}~\bibnamefont
  {Bruno}}\ and\ \bibinfo {author} {\bibfnamefont {C.}~\bibnamefont
  {Chappert}},\ }\href {https://doi.org/10.1103/PhysRevB.46.261} {\bibfield
  {journal} {\bibinfo  {journal} {Phys. Rev. B}\ }\textbf {\bibinfo {volume}
  {46}},\ \bibinfo {pages} {261} (\bibinfo {year} {1992})}\BibitemShut
  {NoStop}%
\bibitem [{\citenamefont {Wang}\ \emph {et~al.}(2017)\citenamefont {Wang},
  \citenamefont {Chang},\ and\ \citenamefont {Zhou}}]{BiTeIsingleimps}%
  \BibitemOpen
  \bibfield  {author} {\bibinfo {author} {\bibfnamefont {S.-X.}\ \bibnamefont
  {Wang}}, \bibinfo {author} {\bibfnamefont {H.-R.}\ \bibnamefont {Chang}},\
  and\ \bibinfo {author} {\bibfnamefont {J.}~\bibnamefont {Zhou}},\ }\href
  {https://doi.org/10.1103/PhysRevB.96.115204} {\bibfield  {journal} {\bibinfo
  {journal} {Phys. Rev. B}\ }\textbf {\bibinfo {volume} {96}},\ \bibinfo
  {pages} {115204} (\bibinfo {year} {2017})}\BibitemShut {NoStop}%
\bibitem [{\citenamefont {Baibich}\ \emph {et~al.}(1988)\citenamefont
  {Baibich}, \citenamefont {Broto}, \citenamefont {Fert}, \citenamefont
  {Van~Dau}, \citenamefont {Petroff}, \citenamefont {Etienne}, \citenamefont
  {Creuzet}, \citenamefont {Friederich},\ and\ \citenamefont
  {Chazelas}}]{mangetores}%
  \BibitemOpen
  \bibfield  {author} {\bibinfo {author} {\bibfnamefont {M.~N.}\ \bibnamefont
  {Baibich}}, \bibinfo {author} {\bibfnamefont {J.~M.}\ \bibnamefont {Broto}},
  \bibinfo {author} {\bibfnamefont {A.}~\bibnamefont {Fert}}, \bibinfo {author}
  {\bibfnamefont {F.~N.}\ \bibnamefont {Van~Dau}}, \bibinfo {author}
  {\bibfnamefont {F.}~\bibnamefont {Petroff}}, \bibinfo {author} {\bibfnamefont
  {P.}~\bibnamefont {Etienne}}, \bibinfo {author} {\bibfnamefont
  {G.}~\bibnamefont {Creuzet}}, \bibinfo {author} {\bibfnamefont
  {A.}~\bibnamefont {Friederich}},\ and\ \bibinfo {author} {\bibfnamefont
  {J.}~\bibnamefont {Chazelas}},\ }\href
  {https://doi.org/10.1103/PhysRevLett.61.2472} {\bibfield  {journal} {\bibinfo
   {journal} {Phys. Rev. Lett.}\ }\textbf {\bibinfo {volume} {61}},\ \bibinfo
  {pages} {2472} (\bibinfo {year} {1988})}\BibitemShut {NoStop}%
\bibitem [{\citenamefont {Grigoriev}\ \emph {et~al.}(2010)\citenamefont
  {Grigoriev}, \citenamefont {Lott}, \citenamefont {Chetverikov}, \citenamefont
  {Gr\"unwald}, \citenamefont {Ward},\ and\ \citenamefont
  {Schreyer}}]{neutorn}%
  \BibitemOpen
  \bibfield  {author} {\bibinfo {author} {\bibfnamefont {S.~V.}\ \bibnamefont
  {Grigoriev}}, \bibinfo {author} {\bibfnamefont {D.}~\bibnamefont {Lott}},
  \bibinfo {author} {\bibfnamefont {Y.~O.}\ \bibnamefont {Chetverikov}},
  \bibinfo {author} {\bibfnamefont {A.~T.~D.}\ \bibnamefont {Gr\"unwald}},
  \bibinfo {author} {\bibfnamefont {R.~C.~C.}\ \bibnamefont {Ward}},\ and\
  \bibinfo {author} {\bibfnamefont {A.}~\bibnamefont {Schreyer}},\ }\href
  {https://doi.org/10.1103/PhysRevB.82.195432} {\bibfield  {journal} {\bibinfo
  {journal} {Phys. Rev. B}\ }\textbf {\bibinfo {volume} {82}},\ \bibinfo
  {pages} {195432} (\bibinfo {year} {2010})}\BibitemShut {NoStop}%
\bibitem [{\citenamefont {Khodadadi}\ \emph {et~al.}(2017)\citenamefont
  {Khodadadi}, \citenamefont {Mohammadi}, \citenamefont {Jones}, \citenamefont
  {Srivastava}, \citenamefont {Mewes}, \citenamefont {Mewes},\ and\
  \citenamefont {Kaiser}}]{FerrRess}%
  \BibitemOpen
  \bibfield  {author} {\bibinfo {author} {\bibfnamefont {B.}~\bibnamefont
  {Khodadadi}}, \bibinfo {author} {\bibfnamefont {J.~B.}\ \bibnamefont
  {Mohammadi}}, \bibinfo {author} {\bibfnamefont {J.~M.}\ \bibnamefont
  {Jones}}, \bibinfo {author} {\bibfnamefont {A.}~\bibnamefont {Srivastava}},
  \bibinfo {author} {\bibfnamefont {C.}~\bibnamefont {Mewes}}, \bibinfo
  {author} {\bibfnamefont {T.}~\bibnamefont {Mewes}},\ and\ \bibinfo {author}
  {\bibfnamefont {C.}~\bibnamefont {Kaiser}},\ }\href
  {https://doi.org/10.1103/PhysRevApplied.8.014024} {\bibfield  {journal}
  {\bibinfo  {journal} {Phys. Rev. Appl.}\ }\textbf {\bibinfo {volume} {8}},\
  \bibinfo {pages} {014024} (\bibinfo {year} {2017})}\BibitemShut {NoStop}%
\bibitem [{\citenamefont {Ishihara}\ \emph {et~al.}(2023)\citenamefont
  {Ishihara}, \citenamefont {Mori}, \citenamefont {Suzuki}, \citenamefont
  {Sato}, \citenamefont {Morita}, \citenamefont {Kohda}, \citenamefont {Ohno},\
  and\ \citenamefont {Miyajima}}]{magopticalkerr}%
  \BibitemOpen
  \bibfield  {author} {\bibinfo {author} {\bibfnamefont {J.}~\bibnamefont
  {Ishihara}}, \bibinfo {author} {\bibfnamefont {T.}~\bibnamefont {Mori}},
  \bibinfo {author} {\bibfnamefont {T.}~\bibnamefont {Suzuki}}, \bibinfo
  {author} {\bibfnamefont {S.}~\bibnamefont {Sato}}, \bibinfo {author}
  {\bibfnamefont {K.}~\bibnamefont {Morita}}, \bibinfo {author} {\bibfnamefont
  {M.}~\bibnamefont {Kohda}}, \bibinfo {author} {\bibfnamefont
  {Y.}~\bibnamefont {Ohno}},\ and\ \bibinfo {author} {\bibfnamefont
  {K.}~\bibnamefont {Miyajima}},\ }\href
  {https://doi.org/10.1103/PhysRevLett.130.126701} {\bibfield  {journal}
  {\bibinfo  {journal} {Phys. Rev. Lett.}\ }\textbf {\bibinfo {volume} {130}},\
  \bibinfo {pages} {126701} (\bibinfo {year} {2023})}\BibitemShut {NoStop}%
\bibitem [{\citenamefont {St{\"o}hr}\ and\ \citenamefont
  {Siegmann}(2006)}]{book-wedge}%
  \BibitemOpen
  \bibfield  {author} {\bibinfo {author} {\bibfnamefont {J.}~\bibnamefont
  {St{\"o}hr}}\ and\ \bibinfo {author} {\bibfnamefont {H.}~\bibnamefont
  {Siegmann}},\ }\href
  {https://link.springer.com/book/10.1007/978-3-540-30283-4} {\emph {\bibinfo
  {title} {Magnetism: From Fundamentals to Nanoscale Dynamics}}},\ Springer
  Series in Solid-State Sciences\ (\bibinfo  {publisher} {Springer},\ \bibinfo
  {address} {Berlin, Heidelberg\;},\ \bibinfo {year} {2006})\BibitemShut
  {NoStop}%
\bibitem [{\citenamefont {Iba\~nez Azpiroz}\ \emph {et~al.}(2013)\citenamefont
  {Iba\~nez Azpiroz}, \citenamefont {Bergara}, \citenamefont {Sherman},\ and\
  \citenamefont {Eiguren}}]{flip}%
  \BibitemOpen
  \bibfield  {author} {\bibinfo {author} {\bibfnamefont {J.}~\bibnamefont
  {Iba\~nez Azpiroz}}, \bibinfo {author} {\bibfnamefont {A.}~\bibnamefont
  {Bergara}}, \bibinfo {author} {\bibfnamefont {E.~Y.}\ \bibnamefont
  {Sherman}},\ and\ \bibinfo {author} {\bibfnamefont {A.}~\bibnamefont
  {Eiguren}},\ }\href {https://doi.org/10.1103/PhysRevB.88.125404} {\bibfield
  {journal} {\bibinfo  {journal} {Phys. Rev. B}\ }\textbf {\bibinfo {volume}
  {88}},\ \bibinfo {pages} {125404} (\bibinfo {year} {2013})}\BibitemShut
  {NoStop}%
\end{thebibliography}%
\end{document}